\magnification1200

\rightline{ King's College-MTH-98-20}
\rightline{hepth/9805055}
\vskip 1cm
\centerline {\bf{{Introduction to Rigid Supersymmetric
Theories}}${\ }^{\ddag}$}
\vskip 1cm
\centerline{\bf P.C. West }
\vskip .2cm
\centerline{Department of Mathematics}
\centerline{King's College, London, UK}

\vskip 1cm
\leftline{\sl Abstract}
 In these lectures we discuss the supersymmetry algebra and its
irreducible representations. We construct the theories of
rigid supersymmetry and gave their superspace formulations. The
perturbative quantum properties of the extended supersymmetric
theories are derived, including the superconformal invariance of a
large class of these theories as well as  the
chiral effective action for
$N=2$ Yang-Mills theory. The superconformal transformations in four
dimensional superspace are derived and encoded into one superconformal
Killing superfield. It is also shown that  the anomalous dimensions of
chiral operators in a superconformal quantum field
 are related to their $R$ weight.
\par
 Some of this material
follows the  book of  reference [0] by the author. Certain
chapters of this book are reproduced, however,  in other sections  the
reader is referred to the relevant parts of reference [0].
In this review, chapter 6 on superconformal theories and
three sections of chapter 5 on flat directions, non-holomorphicity and
the chiral effective action for
$N=2$ Yang-Mills theory are new material.
\par
 The aim of the lectures is to provide the reader with
the material required to understand more recent  developments in the
non-perturbative properties  of quantum extended supersymmetric
theories.

\vskip2cm

${\ }^{\ddag}$
This material is based on lectures presented at   the
Nato conference on "Confinement, Duality, and Nonperturbative Aspects of QCD",
 the Issac Newton Institute, Cambridge, UK
and at the  TASI 1998 Summer School,  Boulder, Colorado, USA.

\vfill\eject

 {\centerline {\bf Contents}}
\medskip
{\bf 1. The Supersymmetry Algebra  }
\medskip
{\bf 2. Models of Rigid  Supersymmetry}

2.1 The Wess-Zumino Model

2.2 The $N=1$ Yang-Mills  Theory

2.3 The Extended Theories
\medskip
{\bf 3 The Irreducible Representations of the Supersymmetry }
\medskip
{\bf 4 Superspace }

4.1 Construction of Superspace

4.2 Superspace Formulations of Rigid Supersymmetric
Theories
\medskip
{\bf 5 Quantum Properties of Supersymmetric Models}

5.1 Super-Feynman Rules and the Non-renormalisation Theorem

5.2 Flat Directions

5.3 Non-holomorphicity

5.4 Perturbative Quantum Properties of Extended Theories of
Supersymmetry

5.5  The $N=2$  Chiral Effective Action
\medskip
{\bf 6 Superconformal Theories}

6.1  The Geometry of Superconformal Transformations.

6.2 Chiral Operators and Fixed Points

\medskip
{\bf References}

\bigskip
{\centerline {\bf Lecture 1. The Supersymmetry Algebra  } }
\medskip
This section is identical to chapter 2 of reference [0].
The equation numbers are keep the same as in this book. I
thank  World Scientific Publishing for their
permission to reproduced this material.
\par
In the 1960's, with the growing awareness of the significance of
internal symmetries such as $SU(2)$ and larger groups, physicists
attempted to
 find a symmetry which would combine in a non-trivial way the
space-time Poincar\'e group with an internal symmetry group. After
much effort it was shown that such an attempt was impossible within
the context of a Lie group. Coleman and Mandula$^4$ showed on very
general assumptions that any Lie group which contained the
Poincar\'e group $P$, whose generators $P_a$ and $J_{ab}$ satisfy the
 relations
$$\eqalignno{[P_a,P_b]&=0\cr
[P_a,J_{bc}]&=(\eta_{ab}P_c-\eta_{ac}P_b)\cr
[J_{ab},J_{cd}]&=-(\eta_{ac}J_{bd}+\eta_{bd}J_{ac}-\eta_{ad}J_{bc}-\eta_{bc}J_
{ad})&(2.1)}$$
and an internal symmetry group $G$ with generators $T_s$ such that
$$[T_r,T_s]=f_{rst}T_t\eqno(2.2)$$
must be a direct product of $P$ and $G$; or in other words
$$[P_a,T_s]=0=[J_{ab},T_s]\eqno(2.3)$$
They also showed that $G$ must be of the form of a semisimple group
with additional $U(1)$ groups.

It is worthwhile to make some remarks concerning the status of this
no-go theorem. Clearly there are Lie groups that contain
the Poincar\'e group and internal symmetry groups in a non-trivial
manner; however, the theorem states that these groups lead to trivial
physics. Consider, for example, two-body scattering; once
we have imposed conservation of angular momentum and momentum the
scattering angle is the only unknown quantity. If there were a Lie
group that had a non-trivial mixing with the Poincar\'e group then
there would be further generators associated with
space-time. The  resulting conservation laws will
further constrain, for example,  two-body scattering, and so the
scattering angle can only take on  discrete values. However, the
scattering process is expected to be  analytic in  the scattering
angle,
$\theta$, and hence we must  conclude  that the process does not
depend on $\theta$ at all.

Essentially the theorem shows that if one used a Lie group that
contained an internal group which mixed in a non-trivial manner with
the Poincar\'e group then the S-matrix for all processes would be zero.
The theorem assumes among other things, that the S-matrix exists
and is non-trivial, the vacuum is non-degenerate and that there are
no   massless particles. It is important to realise that the
theorem only applies to symmetries that act on S-matrix elements and
not on all the other many symmetries that occur in quantum field
theory. Indeed it is not uncommon to find examples of the latter
symmetries. Of course, no-go theorems are only as strong as the
assumptions required to prove them.

In a remarkable paper Gelfand and Likhtman [1] showed that provided
one generalised the concept of a Lie group one could indeed find a
symmetry that included the Poincar\'e group and an internal symmetry
group in a non-trivial way. In this section we will discuss this
approach to the supersymmetry group; having adopted a more general
notion of a group, we will show that one is led, with the aid of
the Coleman-Mandula theorem, and a few assumptions, to the known
supersymmetry group. Since the structure of a Lie group, at least in
some local region of the identity, is determined entirely by its Lie
algebra it is necessary to adopt a more general notion than a Lie
algebra. The vital step in discovering the supersymmetry algebra
is to introduce generators $Q_\alpha^i$, which satisfy
anti-commutation relations, i.e.
$$\eqalignno{\{Q_\alpha^i,Q_\beta^j\}&=Q_\alpha^iQ_\beta^j
+Q_\beta^jQ_\alpha^i\cr
&=\ {\rm some\ other\ generator}&(2.4)}$$
The significance of the $i$ and $\alpha$ indices will become
apparent   shortly. Let us therefore assume that the supersymmetry
group involves  generators $P_a,\ J_{ab},\ T_s$ and possibly some
other  generators which satisfy commutation relations, as well as the
generators $Q_\alpha^i\ (i=1,2,\dots,N)$. We will call the former
generators which satisfy Eqs. (2.1), (2.2) and (2.3) to be even and
those satisfying Eq. (2.4) to be odd generators.

Having let the genie out of the bottle we promptly replace the
stopper and demand that the supersymmetry algebras have a $Z_2$
graded   structure. This simply means that the even and odd generators
must  satisfy the rules:

$$\eqalignno{
[{\rm even},{\rm even}]&={\rm even}\cr
\{{\rm odd},{\rm odd}\}&={\rm even}\cr
[{\rm even},{\rm odd}]&={\rm odd}&(2.5)}$$
We must still have the relations
$$[P_a,T_s]=0=[J_{ab},T_s]\eqno(2.6)$$
since the even (bosonic) subgroup  must obey the Coleman-Mandula
theorem.

Let us now investigate the commutator between $J_{ab}$ and
$Q_\alpha^i$. As a result of Eq.(2.5) it must be of the form
$$[Q_\alpha^i,J_{ab}]=(b_{ab})_\alpha^\beta Q_\beta^i\eqno(2.7)$$
since by definition the $Q_\alpha^i$ are the only odd generators. We
take the $\alpha$ indices to be those rotated by $J_{ab}$. As in a
Lie  algebra we have some generalised Jacobi identities. If we
denote   an even generator by
$B$ and an odd generator by $F$ we find that
$$\eqalignno{
\big[[B_1,B_2],B_3\big]+\big[[B_3,B_1],B_2\big]+\big[[B_2,B_3],B_1\big]&=0\cr
\big[[B_1,B_2],F_3\big]+\big[[F_3,B_1],B_2\big]+\big[[B_2,F_3],B_1\big]&=0\cr
\{[B_1,F_2],F_3\}+\{[B_1,F_3],F_2\}+[\{F_2,F_3\},B_1]&=0\cr
[\{F_1,F_2\},F_3]+[\{F_1,F_3\},F_2]+[\{F_2,F_3\},F_1]&=0&(2.8)}$$
The reader may verify, by expanding each bracket, that these
relations  are indeed identically true.

The identity
$$\big[[J_{ab},J_{cd}],Q_\alpha^i\big]+\big[[Q_\alpha^i,J_{ab}],J_{cd}\big]+\big
[[J_{cd},Q_\alpha^i],J_{ab}\big]=0\eqno(2.9)$$
upon use of Eq. (2.7) implies that
$$[b_{ab},b_{cd}]_\alpha^\beta=-\eta_{ac}(b_{bd})_\alpha^\beta-\eta_{bd}(b_{ac
})_\alpha^\beta+\eta_{ad}(b_{bc})_\alpha^\beta+\eta_{bc}
(b_{ad})_\alpha^\beta\eqno(2.10)$$
This means that the $(b_{cd})_\alpha^\beta$ form a representation of
the Lorentz algebra or in other words the $Q_\alpha^i$ carry a
representation of the Lorentz group. We will select $Q_\alpha^i$ to
be  in the
$(0,{1\over2})\oplus({1\over2},
0)$ representation of the Lorentz group, i.e.
$$[Q_\alpha^i,J_{ab}]={1\over2}(\sigma_{ab})_\alpha^\beta
Q_\beta^i\eqno(2.11)$$
We can choose $Q_\alpha^i$ to be a Majorana spinor, i.e.
$$Q_\alpha^i=C_{\alpha\beta}\bar Q^{\beta i}\eqno(2.12)$$
where $C_{\alpha\beta}=-C_{\beta\alpha}$ is the charge conjugation
matrix (see
appendix A of [0]). This does not represent a loss of generality
since,  if the algebra admits complex conjugation as an involution we
can   always redefine the supercharges so as to satisfy (2.12) (see
Note 1 at the end of this chapter).

The above calculation reflects the more general result that the
$Q_\alpha^i$
must belong to a realization of the even (bosonic) subalgebras of the
supersymmetry
group. This is a simple consequence of demanding that the algebra
be $Z_2$ graded.
The commutator of any even generator $B_1$, with $Q_\alpha^i$ is of
the form
$$[Q_\alpha^i,B_1]=(h_1)_{\alpha j}^{i\beta}Q_\beta^j\eqno(2.13)$$
The generalised Jacobi identity
$$\big[[Q_\alpha^i,B_1],B_2\big]+\big[[B_1,B_2],Q_\alpha^i\big]+\big[[B_2,Q_
\alpha^i],B_1\big]=0\eqno(2.14)$$
implies that
$$[h_1,h_2]_{\alpha
j}^{i\beta}Q_\beta^j=\big[Q_\alpha^i[B_1,B_2]\big]\eqno(2.15)$$
or in other words the matrices $h$ represent the Lie algebra of the
even generators.

The above remarks imply that
$$[Q_\alpha^i,T_r]=(l_r)^i_jQ_\alpha^j+(t_r)^i_j(i\gamma_5)_\alpha^\beta
Q_\beta^j\eqno(2.16)$$
where $(l_r)^i_j+i\gamma_5(t_r)^i_j$ represent the Lie algebra of
the internal
symmetry group. This results from the fact that
$\delta_\beta^\alpha$   and $(\gamma_5)_\alpha^\beta$ are the only
invariant tensors which  are  scalar and pseudoscalar.

The remaining odd-even commutator is $[Q_\alpha^i,P_a]$. A
possibility  that is allowed by the generalised Jacobi identities
that  involve the internal symmetry
group and the Lorentz group is
$$[Q_\alpha^i,P_a]=c(\gamma_a)_\alpha^\beta Q_\beta^i\eqno(2.17)$$
However, the $\big[[Q_\alpha^i,P_a],P_b\big]+\dots$ identity implies
that the constant $c=0$, i.e.
$$[Q_\alpha^i,P_a]=0\eqno(2.18)$$
More generally we could have considered
$(c\gamma_a+d\gamma_a\gamma_5)Q$, on the
right-hand side of (2.17), however, then the above Jacobi identity
and  the Majorana
condition imply that $c=d=0$. (See Note 2 at the end of this
chapter).  Let us
finally consider the $\{Q_\alpha^i,Q_\beta^j\}$ anticommutator. This
object must
be composed of even generators and must be symmetric under
interchange  of $\alpha
\ \leftrightarrow\ \beta$ and $i\ \leftrightarrow\ j$. The even
generators are those of the Poincar\'e group, the internal symmetry
group and other even generators which, from the Coleman-Mandula
theorem, commute with the Poincar\'e group, i.e.
they are scalar and pseudoscalar. Hence the most general possibility
is of the  form
$$\{Q_\alpha^i,Q_\beta^j\}=r(\gamma^aC)_{\alpha\beta}P_a\delta^{ij}+s(\sigma^{
ab}C)_{\alpha\beta}J_{ab}\delta^{ij}+C_{\alpha\beta}U^{ij}+(\gamma_5C)_{\alpha
\beta}V^{ij}\eqno(2.19)$$
In this equation $U^{ij}$ and $V^{ij}$ are new generators which we
will discuss further below. We have not included a
$(\gamma^b\gamma_5C)_{\alpha\beta}L_b^{ij}$ term  as the
$(Q,Q,J_{ab})$ Jacobi identity implies that $L_b^{ij}$ mixes
nontrivially with
the Poincar\'e group and so is excluded by the no-go theorem.

The fact that we have only used numerically invariant tensors under
the Poincar\'e
group is a consequence of the generalised Jacobi identities between
two odd and
one even generators.

To illustrate the argument more clearly, let us temporarily
specialise  to the  case $N=1$ where there is only one supercharge
$Q_\alpha$. Equation (2.19) then
reads
$$\{Q_\alpha,Q_\beta\}=r(\gamma^aC)_{\alpha\beta}P_a+s(\sigma^{ab}C)_{\alpha
\beta}J_{ab}.$$
Using the Jacobi identity
$$\{[P_a,Q_\alpha],Q_\beta\}+\{[P_a,Q_\beta],Q_\alpha\}+[\{Q_\alpha,Q_\beta\},
P_a]=0,$$
we find that
$$0=s(\sigma^{cd}C)_{\alpha\beta}[J_{cd},P_a]
=s(\sigma^{cd}C)_{\alpha\beta}(-
\eta_{ac}P_d+\eta_{ad}P_c),$$
and, consequently, $s=0$. We are free to scale the generator $P_a$
in   order to
bring $r=2$.

Let us now consider the commutator of the generator of the internal
group and
the supercharge. For only one supercharge, Eq. (2.16) reduces to
$$[Q_\alpha,T_r]=l_rQ_\alpha+i(\gamma_5)^\beta_\alpha t_rQ_\beta.$$
Taking the adjoint of this equation, multiplying by $(i\gamma^0)$ and
using the
definition of the Dirac conjugate given in appendix A of reference
[0], we find that
$$[\bar Q^\alpha,T_r]=l^*_r\bar Q^\alpha+\bar
Q^\beta(it^*_r)(\gamma_5)_\beta^
\alpha.$$
Multiplying by $C_{\gamma\alpha}$ and using Eq. (2.12), we arrive at
the equation
$$[Q_\alpha,T_r]=l^*_rQ_\alpha+it^*_r(\gamma_5)_\alpha^\beta Q_\beta.$$
Comparing this equation with the one we started from, we therefore
conclude that
$$l^*_r=l_r,\qquad t^*_r=t_r.$$
The Jacobi identity
$$[\{Q_\alpha,Q_\beta\},T_r]+\big[[T_r,Q_\alpha],Q_\beta\big]
+\big[[T_r,Q_\beta
],Q_\alpha\big]=0$$
results in the equation
$$\eqalignno{
[0+\big(l_r\delta_\alpha^\gamma+it_r(\gamma_5)\alpha^\gamma\big)
2(\gamma_aC)_{
\gamma\beta}P_a]+(\alpha\ \leftrightarrow\ \beta)&=\cr
2P_a\{l_r(\gamma_aC)_{\alpha\beta}+it_r(\gamma_5\gamma_aC)_{\alpha\beta}\}+(
\alpha\ \leftrightarrow\ \beta)&=0.}$$
Since $(\gamma_aC)_{\alpha\beta}$ and
$(\gamma_5\gamma_aC)_{\alpha\beta}$ are
symmetric and antisymmetric in $\alpha\beta$ respectively, we
conclude  that $l_r=0$ but $t_r$ has no constraint placed on it.
Consequently, we  find that we have
only one internal generator $R$ and we may scale it such that
$$[Q_\alpha,R]=i(\gamma_5)_\alpha^\beta Q_\beta.$$
The $N=1$ supersymmetry algebra is summarised in Eq. (2.27).

Let us now return to the extend supersymmetry algebra. The even
generators $U^
{ij}=-U^{ji}$ and $V^{ij}=-V^{ji}$ are called central charges [5] and
are often
also denoted by $Z^{ij}$. It is a consequence of the generalised
Jacobi  identities
\big($(Q,Q,Q)$ and $(Q,Q,Z)$\big) that they commute with all other
generators
including themselves, i.e.
$$[U^{ij},{\rm anything}]=0=[V^{ij},{\rm anything}]\eqno(2.20)$$
We note that the Coleman-Mandula theorem allowed a semi-simple group
plus $U(1)$ factors. The details of the calculation are given in  note
5 at the end of the chapter. Their role in supersymmetric
theories will emerge in later  chapters.

In general, we should write, on the right-hand side of (2.19),
$(\gamma^aC)_{
\alpha\beta}\omega^{ij}P_a+\dots$, where $\omega^{ij}$ is an
arbitrary  real symmetric
matrix. However, one can show that it is possible to redefine (rotate
and rescale)
the supercharges, whilst preserving the Majorana condition, in such a
way as to
bring $\omega^{ij}$ to the form $\omega^{ij}=r\delta^{ij}$ (see Note
3  at the
end of this chapter). The $[P_a,\{Q_\alpha^i,Q_\beta^j\}]+\dots=0$
identity implies
that $s=0$ and we can normalise $P_a$ by setting $r=2$ yielding the
final result
$$\{Q_\alpha^i,Q_\beta^j\}=2(\gamma_aC)_{\alpha\beta}\delta^{ij}P^a+C_{\alpha
\beta}U^{ij}+(\gamma_5C)_{\alpha\beta}V^{ij}\eqno(2.21)$$
In any case $r$ and $s$ have different dimensions and so it would
require the
introduction of a dimensional parameter in order that they were both
non-zero.

Had we chosen another irreducible Lorentz representation for
$Q_\alpha^i$ other
than $(j+{1\over2},j)\oplus(j,j+{1\over2})$ we would not have been
able to put
$P_a$, i.e. a $({1\over2},{1\over2})$ representation, on the
right-hand side of
Eq. (2.21). The simplest choice is $(0,{1\over2})\oplus({1\over2},0)$.
In fact
this is the only possible choice (see Note 4).

Finally, we must discuss the constraints placed on the internal
symmetry group
by the generalised Jacobi identity. This discussion is complicated by
the particular
way the Majorana constraint of Eq. (2.12) is written. A two-component
version of this constraint is
$$\bar Q_{\dot Ai}=(Q_A^i)^*;\quad A,\dot A=1,2\eqno(2.22)$$
(see A of  [0] for two-component notation). Equation (2.19) and
(2.16) then become
$$\eqalignno{
\{Q_A^i,\bar Q_{\dot Bj}\}&=-2i(\sigma^a)_{A\dot B}\delta^i_jP_a\cr
\{Q_A^i,Q_B^j\}&=\varepsilon_{AB}(U^{ij}+iV^{ij})\cr
[Q_A^i,J_{ab}]&=+{1\over2}(\sigma_{ab})_A^BQ_B^i&(2.23)\cr
\noalign{and}
[Q_A^i,T_r]&=(l_r+it_r)^i_jQ_A^j&(2.24)}$$
Taking the complex conjugate of the last equation and using the
Majorana condition
we find that
$$[Q_{\dot Ai},T_r]=Q_{\dot Ak}(U_r^\dagger)^k_i\eqno(2.25)$$
where $(U_r)^i_j=(l_r+it_r)^i_j$. The $(Q,\bar Q,T)$ Jacobi identity
then implies
that $\delta^i_j$ be an invariant tensor of $G$, i.e.
$$U_r+U_r^\dagger=0\eqno(2.26)$$
Hence $U_r$ is an antihermitian matrix and so represents the
generators of the
unitary group $U(N)$. However, taking account of the central charge
terms in the
$(Q,Q,T)$ Jacobi identity one finds that there is for every central
charge an
invariant antisymmetric tensor of the internal group and so the
possible internal
symmetry group is further reduced. If there is only one central
charge, the internal
group is $Sp(N)$ while if there are no central charges it is $U(N)$.

To summarise, once we have adopted the rule that the algebra be $Z_2$
graded and
contain the Poincar\'e group and an internal symmetry group then the
generalised
Jacobi identities place very strong constraints on any possible
algebra. In fact,
once one makes the further assumption that $Q_\alpha^i$ are spinors
under the
Lorentz group then the algebra is determined to be of the form of
equations
(2.1), (2.6), (2.11), (2.16), (2.18) and (2.21).

The simplest algebra is for $N=1$ and takes the form
$$\eqalignno{
\{Q_\alpha,Q_\beta\}&=2(\gamma_aC)_{\alpha\beta}P^a\cr
[Q_\alpha,P_a]&=0\cr
[Q_\alpha,J_{cd}]&={1\over2}(\sigma_{cd})_\alpha^\beta Q_\beta\cr
[Q_\alpha,R]&=i(\gamma_5)_\alpha^\beta Q_\beta&(2.27)}$$
as well as the commutation relations of the Poincar\'e group. We note
that there
are no central charges (i.e. $U^{11}=V^{11}=0$), and the internal
symmetry group
becomes just a chiral rotation with generator $R$.

We now wish to prove five of the statements above. This is done here
rather than
in the above text, in order that the main line of argument should not
become
obscured by technical points. These points are best clarified in
two-component
notation.

{\it Note 1:}\quad Suppose we have an algebra that admits a complex
conjugation
as an involution; for the supercharges this means that
$$(Q_A^i)^*=b_i^{\ j} Q_{\dot Aj};\quad(Q_{\dot
Aj})^*=d^j_{\ k}Q_A^k$$  There is no mixing of the Lorentz indices
since $(Q_A^i)^*$ transforms  like $Q_{\dot Ai}$, namely in the
$(0,{1\over2})$ representation of the  Lorentz group,  and not like
$Q_A^i$ which is in the $({1\over2},0)$ representation.  The lowering
of the $i$ index under $*$ is at this point purely a notational
device. Two
successive $*$ operations yield the unit operation and this implies
that
$$(b_i^{\ j})^*d^j_{\ k}=\delta^i_k\eqno(2.28)$$
and in particular that $b_i^{\ j}$ is an invertible matrix. We now
make  the redefinitions
$$\eqalignno{
{Q^\prime}_A^i&=Q_A^i\cr
{Q^\prime}_{\dot Ai}&=b_i^{\ j}Q_{\dot Aj}&(2.29)}$$
Taking the complex conjugate of ${Q^\prime}_{ Ai}$, we find
$$({Q^\prime}_A^i)^*=(Q_A^i)^*=b_i^{\ j}Q_{\dot Aj}={Q^\prime}_{\dot
Aj}$$
while
$$({Q^\prime}_{\dot Ai})^*=(b_i^{\ j})^*(Q_{\dot
Aj})^*=(b_i^{\ j})^*d^j_{\ k}Q_A^k=Q_A^i\eqno(2.30)$$
using Eq. (2.28).

Thus the ${Q^\prime}_A^i$ satisfy the Majorana condition, as
required.  If the
$Q$'s do not initially satisfy the Majorana condition, we may simply
redefine
them so that they do.

{\it Note 2:}\quad Suppose the $[Q_A,P_a]$ commutator were of the
form
$$[Q_A,P_a]=e(\sigma_a)_{A\dot B}Q^{\dot B}\eqno(2.31)$$
where $e$ is a complex number and for simplicity we have suppressed
the $i$ index.
Taking the complex conjugate (see A of reference [0]), we find that
$$[Q_{\dot A},P_a]=-e^*(\sigma_a)_{B\dot A}Q^B\eqno(2.32)$$
Consideration of the $\big[[Q_A,P_a],P_b\big]+\dots=0$ Jacobi
identity  yields
the result
$$-|e|^2(\sigma_a)_{A\dot B}(\sigma^b)^{C\dot B}-(a\ \leftrightarrow
b)=0\eqno(2.33)$$
Consequently $e=0$ and we recover the result
$$[Q_A,P_a]=0\eqno(2.34)$$

{\it Note 3:}\quad The most general form of the $Q^{Ai},Q^{\dot B}_j$
anticommutator
is
$$\{Q^{Ai},Q^{\dot B}_j\}=-2iU^i_j(\sigma^m)^{A\dot B}P_m+{\rm terms\
involving
\ other\ Dirac\ matrices}\eqno(2.35)$$
Taking the complex conjugate of this equation and comparing it with
itself, we
find that $U$ is a Hermitian matrix
$$(U^i_j)^*=U^j_i\eqno(2.36)$$
We now make a field redefinition of the supercharge
$${Q^\prime}^{Ai}=B^i_jQ^{Aj}\eqno(2.37)$$
and its complex conjugate
$${Q^\prime}^{\dot A}_i=(B^i_j)^*Q^{\dot A}_j\eqno(2.38)$$
Upon making this redefinition in Eq. (2.35), the $U$ matrix becomes
replaced by
$${U^\prime}^i_j=B^i_kU^k_l(B^j_l)^*\quad{\rm or}\quad
U^\prime=BUB^\dagger\eqno(2.39)$$
Since $U$ is a Hermitian matrix, we may diagonalise it in the form
$c_i\delta^
i_j$ using a unitarity matrix $B$. We note that this preserves the
Majorana
condition on $Q^{Ai}$. Finally, we may scale $Q^i\rightarrow(1/\sqrt
c^i)Q^i$
to bring $U$ to the form $U=d_i\delta^i_j$, where $d_i=\pm1$. In
fact,  taking
$A=B=1$ and $i=j=k$, we realise that the right-hand side of Eq.
(2.35)  is a positive
definite operator and since the energy $-iP_0$ is assumed positive
definite, we
can only find $d_i=+1$. The final result is
$$\{Q^{Ai},Q^{\dot B}_j\}=-2i\delta^i_j(\sigma^m)^{A\dot
B}P_m\eqno(2.40)$$

{\it Note 4:}\quad Let us suppose that the supercharge $Q$ contains an
irreducible
representation of the Lorentz group other than
$(0,{1\over2})\oplus({1\over2},
0)$, say, the representation $Q_{A_1\dots A_n,\dot B_1\dots\dot B_m}$
where the
$A$ and $B$ indices are understood to be separately symmetrised and
$n+m$ is odd
in order that $Q$ is odd and $n+m>1$. By projecting the
$\{Q,Q^\dagger\}$ anticommutator
we may find the anti-commutator involving $Q_{A_1\dots A_n,\dot
B_1\dots\dot B
_m}$ and its hermitian conjugate. Let us consider in particular the
anticommutator
involving $Q=Q_{11\dots1,\dot 1\dot 1\dots \dot 1}$, this must result
in an object of spin $n +m>1$. However, by the Coleman-Mandula no-go
theorem no such generator  can occur
in the algebra and so the anticommutator must vanish, i.e.
$QQ^\dagger+Q^\dagger
Q=0$.

Assuming the space on which $Q$ acts has a positive definite norm, one
such example
being the space of on-shell states, we must conclude that $Q$
vanishes. However
if $Q_{11\dots1,\dot 1\dot 1\dots \dot 1}$ vanishes, so must
$Q_{A_1\dots A_n,\dot  B_1\dots\dot
B_m}$ by its Lorentz properties, and we are left only with the
$(0,{1\over2})
\oplus({1\over2},0)$ representation.

{\it Note 5:}\quad We now return to the proof of equation (2.20).
Using the (Q,Q,Z) Jacobi identity it is straightforward to show that
the supercharges $Q$ commute with the central charges $Z$. The
(Q,Q,U)  Jacobi identity then implies that the central charges
commute with themselves. Finally, one considers the $(Q,Q,T_r)$ Jacobi
identity. This relation shows that the commutator of $T_r$ and $Z$
takes the generic form $[T_r,\ Z]= \dots Z$. However, the generators
$T_r$ and $Z$ form the internal symmetry group of the supersymmetry
algebra and from the no-go theorem we know that this group must be a
semisimple Lie group times
$U(1)$ factors. We recall that a semisimple Lie group is one that has
no normal Abelian subgoups other that the group itself and the identity
element.  As such, we must conclude that $T_r$ and $Z$  commute,
and hence our final result that the central charges commute with all
generators, that is they really are central.

Although the above discussion started with the Poincar\'e group, one
could equally
well have started with the conformal or (anti-) de Sitter groups and
obtained the
superconformal and super (anti)de Sitter algebras. For completeness,
we now list
these algebras. The superconformal algebra which has the generators
$P_n,J_{mn},
D,K_n,A,Q^{\alpha i}$, $S^{\alpha i}$ and the internal symmetry
generators $T_r$ and $A$  is given by the Lorentz group  plus:
$$\eqalignno{
[J_{mn},P_k]&=\eta_{nk}P_m-\eta_{mk}P_n\cr
[J_{mn},K_k]&=\eta_{nk}K_m-\eta_{mk}K_n\cr
[D,P_K]&=-P_K\quad[D,K_K]=+K_K\cr
[P_m,K_n]&=-2J_{mn}+2\eta_{mn}D\quad[K_n,K_m]=0\quad[P_n,P_m]=0\cr
[Q_\alpha^i,J_{mn}]&={1\over2}(\gamma_{mn})_\alpha^\beta
Q_\beta^i\quad[S_\alpha
^i,J_{mn}]={1\over2}(\gamma_{mn})_\alpha^\beta S^{\beta i}\cr
\{Q_\alpha^i,Q_\beta^j\}&=-2(\gamma^nC^{-1})_{\alpha\beta}P_n\delta^{ij}\cr
\{S_\alpha^i,S_\beta^j\}&=+2(\gamma^nC^{-1})_{\alpha\beta}K_n\delta^{ij}\cr
[Q_\alpha^i,D]&={1\over2}Q_\alpha^i\quad[S_\alpha^i,D]=-{1\over2}S_\alpha^i\cr
[Q_\alpha^i,K_n]&=-(\gamma_n)_\alpha^\beta
S_\beta^i\quad[S_\alpha^i,P_n]=(\gamma
_n)_\alpha^\beta Q_\beta^i\cr
[Q_\alpha^i,T_r]&=\big( \delta_\alpha^\beta(\tau_{r_1})^i_j
 +(\gamma_5)_\alpha^
\beta(\tau_{r_2})^i_j\big) Q_\beta^j\cr
[S_\alpha^i,T_r]&=\big(\delta_\alpha^\beta(\tau_{r_1})^i_j-(\gamma_5)_\alpha^
\beta(\tau_{r_2})^i_j\big)Q_\beta^j\cr
[Q_\alpha^i,A]&=-i(\gamma_5)_\alpha^\beta
Q_\beta^i\left({4-N\over4N}\right)\cr
[S_\alpha^i,A]&={4-N\over4N}i(\gamma_5)_\alpha^\beta S_\beta^i\cr
\{Q_\alpha^i,S_\beta^j\}&=-2(C^{-1}_{\alpha\beta})
D\delta^{ij}+(\gamma^{mn}C^{-
1})_{\alpha\beta}J_{mn}\delta^{ij}+4i(\gamma_5C^{-1}_{\alpha\beta})
A\delta^{ij}\cr
&\quad-2(\tau_{r_1})^{ij}(C^{-1})_{\alpha\beta}+\big((\tau_{r_2})^{ij}(\gamma_
5C^{-1})_{\alpha\beta}\big)T_r&(2.41)}$$
The $T_r$ and $A$ generate $U(N)$ and $\tau_1+\gamma_5\tau_2$ are in
the fundamental
representation of $SU(N)$.

The case of $N=4$ is singular and one can have either
$$[Q_\alpha^i,A]=0\quad{\rm
or}\quad[Q_\alpha^i,A]=-i(\gamma_5)_\alpha^\beta Q
_\beta^i$$
and similarly for $S_\alpha^i$ and $A$. One may verify that both
possibilities
are allowed by the $N=4$ Jacobi identities and so form acceptable
superalgebras.

The anti-de Sitter superalgebra has generators $M_{mn},\
T_{ij}=-T_{ji}$ and
$Q^{\alpha i}$, and is given by
$$\eqalignno{
[M_{mn},M_{pq}]&=\eta_{np}M_{mq}+3\ {\rm terms}\cr
[M_{mn},T_{ij}]&=0\quad[Q_\alpha^i,M_{mn}]={1\over2}
(\gamma_{mn})_\alpha
^{\ \beta} Q_\beta^i\cr
[Q_\alpha^i,T^{jk}]&=-2i(\delta^{ij}Q_\alpha^k-\delta^{ik}Q_\alpha^j)\cr
\{Q_\alpha^i,Q_\beta^j\}&=\delta^{ij}(\gamma_{mn}C^{-1})_{\alpha\beta}iM_{mn}+
(C^{-1})_{\alpha\beta}T^{ij}\cr
[T^{ij},T^{kl}]&=-2i(\delta^{jk}T^{il}+3\ {\rm terms})&(2.42)}$$

\bigskip
{\centerline {\bf Lecture  2. Models of Rigid  Supersymmetry} }
\bigskip
{\bf 2.1 The Wess-Zumino Model}
\medskip
This section is identical to chapter 5 of reference [0].
The equation numbers are keep the same as in this book. I
thank  World Scientific Publishing for their
permission to reproduced this material.
\par
The first four-dimensional model in which supersymmetry was
linearly realised
was found by Wess and Zumino$^3$ by studying two-dimensional dual
models$^7$.
In this chapter we rediscover supersymmetry along the lines given in
Chapter 4
and discuss the Wess-Zumino model which is the simplest model of
$N=1$  supersymmetry.

Let us assume that the simplest model possesses one fermion
$\chi_\alpha$ which
is a Majorana spinor, i.e.
$$\chi_\alpha=C_{\alpha\beta}\bar\chi^\beta\eqno(5.1)$$
On shell, that is, when
$$/\!\!\!\partial\chi=\ {\rm interaction}\eqno(5.2)$$
$\chi_\alpha$ has two degrees of freedom or two helicity states.
Applying our
rule concerning equal numbers of fermionic and bosonic degrees of
freedom of
the previous chapter to the on-shell states we find that we must add
two bosonic
degrees of freedom to $\chi_\alpha$ in order to form a realization of
supersymmetry.
These could either be two spin-zero particles or one massless vector
particle which also has two helicity states on-shell. We will
consider  the former possibility
in this section and the latter possibility, which is the $N=1$
Yang-Mills theory,
in the next chapter.

In Chapter 8 we will show that these considerations are indeed correct.
An irreducible representation of $N=1$ supersymmetry can be carried
either by one   parity even
spin-zero state, one parity odd spin-zero state and one Majorana
spin-${1\over2}$, or by one massless spin-one and one Majorana
spin-${1\over2}$. Taking the former possibility we have a Majorana
spinor $\chi_\alpha$ and two
spin-zero states which we will assume to be represented by a scalar
field $A$
and pseudoscalar field $B$. For simplicity we will begin by
constructing the
free theory; the fields $A,\ B,\ \chi_\alpha$ are then subject to
$$\partial^2A=\partial^2B=/\!\!\!\partial\chi=0\eqno(5.3)$$
We now wish to construct the supersymmetry transformations that are
carried by
this irreducible realization of supersymmetry. Since
$\bar\varepsilon^\alpha Q_\alpha$ is dimensionless and $Q_\alpha$ has
mass dimension
$+{1\over2}$, the parameter $\bar\varepsilon^\alpha$ must have
dimension $-{1\over2}$. On grounds
of linearity, dimension, Lorentz invariance and parity we may write
down the following set of transformations:
$$\eqalignno{
\delta A&=\bar\varepsilon QA=\bar\varepsilon\chi\qquad\delta
B=i\bar\varepsilon
\gamma_5\chi\cr
\delta\chi&=/\!\!\!\partial(\alpha A+\beta i\gamma_5B)\varepsilon&(5.4)}$$
where $\alpha$ and $\beta$ are undetermined parameters.

The variation of $A$ is straightforward; however, the appearance of
a derivative
in $\delta\chi$ is the only way to match dimensions once the
transformations
are assumed to be linear. The reader will find no trouble verifying
that these
transformations do leave the set of field equations of Eq.(5.3)
intact.

We can now test whether the $N=1$ supersymmetry algebra of Chapter 2
is represented
by these transformations. The commutator of two supersymmetries on $A$
is given by
$$[\delta_1,\delta_2]A=[\bar\varepsilon_1Q,\bar\varepsilon_2Q]A\eqno(5.5)$$
which, using Eq.(2.27), becomes
$$\eqalignno{
[\delta_1,\delta_2]A&=2\bar\varepsilon_2\gamma^a\varepsilon_1P_aA\cr
&=2\bar\varepsilon_2\gamma^a\varepsilon_1\partial_aA\cr
{\rm since}\kern1in P_a&=\partial_a&(5.6)}$$
On the other hand the transformation laws of Eq.(5.4) imply that
$$\eqalignno{
[\delta_1,\delta_2]A&=\bar\varepsilon_2/\!\!\!\partial(\alpha A+i\gamma_5\beta
B)\varepsilon_1-(1\leftrightarrow2)\cr
&=2\alpha\bar\varepsilon_2/\!\!\!\partial\varepsilon_1A&(5.7)}$$
The term involving $B$ drops out because of the properties of Majorana
spinors
(see appendix A of reference [0]). Provided $\alpha=+1$ this is indeed
the  4-translation required
by the algebra. We therefore set $\alpha=+1$. The calculation for $B$
is similar
and yields $\beta=+1$. For the field $\chi_\alpha$ the commutator of
two supersymmetries
gives the result
$$\eqalignno{
[\delta_{\varepsilon_1},\delta_{\varepsilon_2}]\chi&=/\!\!\!\partial
[\bar\varepsilon_1
\chi+i\gamma_5\bar\varepsilon_1i\gamma_5\chi]\varepsilon_2-(1\leftrightarrow
2)\cr
&=
-{1\over4}\bar\varepsilon_1\gamma^R\varepsilon_2/\!\!\!\partial
[\gamma_R+i\gamma_5
\gamma_Ri\gamma_5]\chi-(1\leftrightarrow2)\cr
&=+{1\over2}\bar\varepsilon_2\gamma^a
\varepsilon_12/\!\!\!\partial\gamma_a
\chi\cr
&=2\bar\varepsilon_2/\!\!\!\partial\varepsilon_1\chi
-\bar\varepsilon_2
\gamma^a\varepsilon_1\gamma_a/\!\!\!\partial\chi&(5.8)}$$
The above calculation makes use of a Fierz rearrangement (see Appendix
A) as well
as the properties of Majorana spinors. However, $\chi_\alpha$ is
subject to its
equation of motion, i.e. $/\!\!\!\partial\chi=0$, implying the final result
$$[\delta_1,\delta_2]\chi=2\bar\varepsilon_2/\!\!\!\partial\varepsilon_1
\chi\eqno(5.9)$$
which is the consequence dictated by the supersymmetry algebra. The
reader will
have no difficulty verifying that the fields $A,\ B$ and
$\chi_\alpha$  and the transformations
$$\eqalignno{
\delta A&=\bar\varepsilon\chi,\quad\delta
B=i\bar\varepsilon\gamma_5\chi\cr
\delta\chi&=/\!\!\!\partial(A+i\gamma_5B)\varepsilon&(5.10)}$$
form a representation of the whole of the supersymmetry algebra {\it
provided}
$A,\ B$ and $\chi_\alpha$ are on-shell (i.e.
$\partial^2A=\partial^2B=/\!\!\!\partial\chi=0$).

We now wish to consider the fields $A,\ B$ and $\chi_\alpha$ when they
are no longer subject to their field equations. The Lagrangian from
which the above

field equations follow is 
$$
L=-{1\over 2}(\partial_\mu A)^2-{1\over 2}(\partial_\mu B)^2
-{1\over 2}{\bar \chi}/ \!\!\!\partial\chi\eqno(5.11)
$$

It is easy to prove that the action $\int d^4xL$ is indeed invariant under the
transformation of Eq.(5.10). This invariance is achieved {\it without
the use
of the field equations}. The trouble with this formulation is that the
fields
$A,\ B$ and $\chi_\alpha$ do not form a realization of the
supersymmetry algebra
when they are no longer subject to their field equations, as the last
term in
Eq.(5.8) demonstrates. It will prove useful to introduce the following
terminology.
We shall refer to an irreducible representation of supersymmetry
carried by
fields which are subject to their equations of motion as an {\it
on-shell
representation}. We shall also refer to a Lagrangian as being {\it
algebraically
on-shell} when it is formed from fields which carry an on-shell
representation,
that is, do not carry a representation of supersymmetry off-shell, and
the
Lagrangian is invariant under these on-shell transformations. The
Lagrangian
of Eq.(5.11) is then an algebraically on-shell Lagrangian.

That $A,\ B$ and $\chi_\alpha$ cannot carry a representation of supersymmetry
off-shell can be seen without any calculation, since these fields do
not satisfy
the rule of equal numbers of fermions and bosons which was given
earlier. Off-shell,
$A$ and $B$ have two degrees of freedom, but $\chi_\alpha$ has four
degrees of
freedom. Clearly, the representations of supersymmetry must change
radically
when enlarged from on-shell to off-shell.

A possible way out of this dilemma would be to add two bosonic fields $F$ and
$G$ which would restore the fermion-boson balance. However, these
additional
fields would have to occur in the Lagrangian so as to give rise to no
on-shell
states. As such, they must occur in the Lagrangian in the form
$+{1\over2}F^2+
{1\over2}G^2$ assuming the free action to be only bilinear in the
fields and
consequently be of mass dimension two. On dimensional grounds their
supersymmetry
transformations must be of the form
$$\delta F=\bar\varepsilon/\!\!\!\partial\chi\quad\delta
G=i\bar\varepsilon\gamma_5/\!\!\!\partial
\chi\eqno(5.12)$$
where we have tacitly assumed that $F$ and $G$ are scalar and
pseudoscalar
respectively. The fields $F$ and $G$ cannot occur in $\delta A$ on
dimensional
grounds, but can occur in $\delta\chi_\alpha$ in the form
$$\delta\chi=[(\mu
F+i\tau\gamma_5G)+/\!\!\!\partial(A+i\gamma_5B)]\varepsilon\eqno(5.13)$$
where $\mu$ and $\tau$ are undetermined parameters.

We note that we can only modify transformation laws in such a way that
on-shell
(i.e., when $F=G=/\!\!\!\partial\chi=\partial^2A=\partial^2B=0$) we regain the
on-shell
transformation laws of Eq.(5.10).

We must now test if these new transformations do form a realization of
the supersymmetry algebra. In fact, straightforward calculation shows
they do, provided $\mu=\tau=+1$. This representation of supersymmetry
involving the fields $A,\ B,\ \chi_\alpha,\ F$ and $G$ was found by
Wess and Zumino$^3$ and we now summarize their result:
$$\eqalignno{
\delta A&=\bar\varepsilon\chi\qquad\delta
B=i\bar\varepsilon\gamma_5\chi\cr
\delta\chi&=[F+i\gamma_5G+/\!\!\!\partial(A+i\gamma_5B)]\varepsilon\cr
\delta
F&=\bar\varepsilon/\!\!\!\partial\chi\qquad\delta G=
i\bar\varepsilon\gamma_5 /\!\!\!\partial
\chi&(5.14)}$$
The action which is invariant under these transformations, is given by
the Lagrangian
$$A=\int d^4x\left\{-{1\over2}(\partial_\mu
A)^2-{1\over2}(\partial_\mu
B)^2-{1\over2}\bar\chi/\!\!\!\partial\chi+{1\over2}F^2+{1\over2}G^2\right\}\
eqno(5.15)$$
As expected the $F$ and $G$ fields occur as squares without
derivatives and so lead to no on-shell states.

The above construction of the Wess-Zumino model is typical of that
for  a general
free supersymmetric theory. We begin with the on-shell states, given
for any
model in Chapter 8, and construct the on-shell transformation laws.
We  can then find the Lagrangian which is invariant without use of the
equations of motion, but contains no auxiliary fields. One then tries
to find a set of auxiliary fields that give an off-shell algebra.
Once  this is done one can find a corresponding off-shell action. How
one
 finds the nonlinear theory from the free theory is
 discussed in the later chapters.

The first of these two steps is always possible; however, there is no
sure way of finding auxiliary fields that are required in all models,
except with a few rare exceptions. This fact is easily seen to be a
consequence of our rule for equal numbers of fermi and bose degrees
of  freedom in any representation of supersymmetry. It is only spin
0's,  when represented by scalars, that have the same number of field
components off-shell as they have on-shell states. For example, a
Majorana spin-${1\over2}$ when represented by a spinor
$\chi_\alpha$ has a jump of 2 degrees of freedom between on and
off-shell and a massless spin-1 boson when represented by a vector
 $A_\mu$ has a jump of 1 degree of freedom. In the latter case it is
important to subtract the one gauge degree of freedom
 from $A_\mu$ thus leaving 3 field components off-shell (see next
chapter). Since the increase in the number of degrees of freedom from
an on-shell state to the off-shell field representing it changes by
different amounts for fermions and bosons, the fermionic-boson
balance  which holds on-shell will not hold off-shell if we only
introduce the fields that describe the on-shell states.
The discrepancy must be made up by fields, like $F$ and $G$, that
lead  to no  on-shell states. These latter type of fields are called
{\it auxiliary fields}. The whole problem of finding representations
of  supersymmetry amounts to finding the auxiliary fields.

Unfortunately, it is not at all easy to find the auxiliary fields.
Although the fermi-bose counting rule gives a guide to the number of
auxiliary fields it does not actually tell you what they are, or how
they transform. In fact, the auxiliary fields are only known for almost
all $N=1$ and 2 supersymmetry theories and for a very few $N=4$
theories and not for the higher $N$ theories. In particular, they are
not known for the $N=8$ supergravity theory.

Theories for which the auxiliary fields are not known can still be
described by a Lagrangian in the same way as the Wess-Zumino theory can
be described without the use of $F$ and $G$, namely, by the so called
algebraically on-shell Lagrangian formulation, which for the
Wess-Zumino theory was given in Eq. (5.11). Such `algebraically
on-shell Lagrangians' are not too difficult to find at least at the
linearized level. As explained in Chapter 8 we can easily find the
relevant on-shell states of the theory. The algebraically on-shell
Lagrangian then consists of writing down the known kinetic terms for
each spin.

Of course, we are really interested in the interacting theories. The
form of the interactions is however often governed by symmetry
principles such as gauge invariance in the above example or general
coordinate invariance in the case of gravity theories. When the form of
the interactions is dictated by a local symmetry there is a
straightforward, although maybe very lengthy way of finding
the nonlinear theory from the linear theory. This method, called
Noether coupling, is described in Chapter 7. In one guise or another
this technique has been used to construct nonlinear `algebraically
on-shell Lagrangians' for all supersymmetric theories.

The reader will now ask himself whether algebraically on-shell
Lagrangians may be good enough. Do we really need the auxiliary fields?
This question will be addressed in the next chapter, but the following
example is a warning  against over-estimating the importance of a
Lagrangian that is invariant under a set of transformations that mix
fermi-bose fields, but do not obey any particular algebra.

Consider the Lagrangian
$$L=-{1\over2}(\partial_\mu
A)^2-{1\over2}\bar\chi/\!\!\!\partial\chi\eqno(5.16)$$
whose corresponding action is invariant under the transformations
$$\partial A=\bar\varepsilon\chi\quad\partial\chi=/\!\!\!\partial
A\varepsilon\eqno(5.17)$$
However, this theory has nothing to do with
supersymmetry. The algebra of transformations of Eq. (5.17) does not
close on or off-shell without generating transformations which,
although invariances of the free theory, can never be generalized to
be invariances of an interacting theory. In fact, the on-shell states
do not even have the correct fermi-bose balance required to form an
irreducible representation of supersymmetry. This example illustrates
the fact that the `algebraically on-shell Lagrangians' rely for their
validity, as supersymmetric theories, on their on-shell algebra.

As a final remark in this section it is worth pointing out that the
problem of finding the representations of any group is a mathematical
question not dependent on any dynamical considerations for its
resolution. Thus the questions of which are physical fields and which
are auxiliary fields is a model-dependent statement.
\bigskip
{\bf 2.2 The $N=1$ Yang-Mills  Theory}
\medskip
This account of the construction of the $N=1$ Yang-Mills
theory in $x$-space follows closely chapter 6 of reference [0].
\bigskip
{\bf 2.3 The Extended Theories}
\medskip
The  $N=2$ Yang-Mills theory and $N=2$ matter are constructed as
well as their most general renormalizable coupling. This account
closely follows chapter 12 of reference [0].
\bigskip
{\centerline {\bf Lecture 3. The Irreducible Representations of
 Supersymmetry }}
\medskip
The first part  of this section  is taken from reference [0]
and we have kept the equation numbers the same as in that
reference.
\medskip
In this chapter we wish to find the irreducible representations of
supersymmetry [11], or, put another way, we want to know what is the
possible particle content of supersymmetric theories. As is well
known  the irreducible representations of the Poincar\'e group are
found by  the Wigner method of induced representations [12]. This
method  consists of finding a representation of a subgroup of the
Poincar\'e  group and boosting it up to a representation of the full
group. In  practice, one adopts the following recipe: we choose a
given momentum
$q^\mu$ which satisfies $q^\mu q_\mu =0$ or $q^\mu q_\mu =-m^2$
depending which  case we are considering. We find the subgroup $H$
which leaves
$q^\mu$  intact and find a representation of $H$  on the
$|q^\mu\rangle$ states.  We then induce this representation to the
whole of the Poincar\'e  group $P$, in the usual way. In this
construction there is a  one-to-one correspondence between points of
$P/H$ and four-momentum  which satisfies $P_\mu P^\mu=0$ or
$P_\mu P^\mu=-m^2$. One can show that the  result
is independent of the
choice of momentum $q^\mu$ one starts  with.

In what follows we will not discuss the irreducible representations
in  general, but only that part applicable to the rest frame, i.e. the
representations of $H$ in the states at rest. We can do this safely
in  the knowledge that once the representation of $H$ on the
rest-frame  states in known then the representation of $P$ is uniquely
given and  that every irreducible representation of the Poincar\'e
group can be  obtained by considering every irreducible representation
of $H$.

In terms of physics the procedure has a simple interpretation,
namely,  the properties of a particle are determined entirely by its
behaviour  in a given frame (i.e. for given $q^\mu$). The general
behaviour is  obtained from the given
$q^\mu$ by boosting either the observer or the frame with momentum
$q^\mu$ to one with arbitrary momentum.

The procedure outlined above for the Poincar\'e group can be
generalised to any group of the form $S\otimes_s T$ where the symbol
$\otimes_s$ denotes the semi-direct product of the groups $S$ and $T$
where $T$ is Abelian. It also applies to the supersymmetry group and
we  shall take it for granted that the above recipe is the correct
procedure and does in fact yield all irreducible representations of
the supersymmetry group.

Let us first consider the massless case $q_\mu q^\mu=0$, for which we
choose the standard momentum $q^\mu_s=(m,0,0,m)$ for our "rest
frame".  We must now find $H$ whose group elements leave
$q^\mu_s=(m,0,0,m)$ intact. Clearly this contains $Q_\alpha^i,\ P_\mu$
and $T_s$, since  these generators all commute with $P_\mu$ and so
rotate the states with
$q^\mu_s$ into themselves. As we will see in the last section one can
not have non-vanishing  central charges for the massless  case.

Under the Lorentz group the action of the generator
${1\over2}\Lambda^{\mu\nu} J_{\mu\nu}$ creates an infinitesimal
transformation $q^\mu\rightarrow\Lambda^
\mu_\nu q^\nu+q^\mu$. Hence $q^\mu_s$ is left invariant provided the
parameters obey the relations
$$\Lambda_{30}=0,\quad\Lambda_{10}+\Lambda_{13}=0,
\quad\Lambda_{20}+\Lambda_{2
3}=0\eqno(8.1)$$
Thus the Lorentz generators in $H$ are
$$T_1=J_{10}+J_{13},\quad T_2=J_{20}+J_{23},\quad J=J_{12}\eqno(8.2)$$
These generators form the algebra
$$\eqalignno{
[T_1,J]&=-T_2\cr
[T_2,J]&=+T_1\cr
[T_1,T_2]&=0&(8.3)}$$
The reader will recognise this to be the Lie algebra of $E_2$, the
group of translations and rotations in a two-dimensional plane.

Now the only unitary representations of $E_2$ which are finite
dimensional have $T_1$ and $T_2$ trivially realised, i.e.
$$T_1|q^\mu_s\rangle=T_2|q^\mu_s\rangle=0
\eqno(8.4)$$
This results from the theorem that all non-trivial unitary
representations of  noncompact groups are infinite dimensional. We
will  assume we require finite-dimensional representations of $H$.

Hence for the Poincar\'e group, in the case of massless particles,
finding representations of $H$ results in finding representations of
$E_2$ and consequently for the generator $J$ alone. We choose our
states so that
$$J|\lambda\rangle=i\lambda|\lambda\rangle
\eqno(8.5)$$
Our generators are antihermitian. In fact, $J$ is the helicity
operator and we select $\lambda$ to be integer or half-integer (i.e.
$J={\underline q}\cdot {\underline  J}/|{\underline q}|$ evaluated at
${\underline q}=(0,0,m)$ where
$J_i=\varepsilon_{ijk}J_{jk};\ i,j=1,2,3)$.

Let us now consider the action of the supercharges $Q_\alpha^i$ on
the  rest-frame states, $|q^\mu_s\rangle$. The calculation is easiest
when  performed using the two-component formulation of the
supersymmetry  algebra of Eq. (2.23). On rest-frame states we find
that
$$\eqalignno{
\{Q^{Ai},Q^{\dot B}_j\}&=-2\delta^i_j(\sigma_\mu)^{A\dot B}q^\mu_s\cr
&=-2\delta^i_j(\sigma_0+\sigma_3)^{A\dot
B}m=+4m\delta^i_j\left(\matrix{
0&0\cr 0&1\cr}\right)^{A\dot B}&(8.6)}$$
In particular these imply the relations
$$\eqalignno{
\{Q^{1i},Q^{\dot 1}_j\}&=0\cr
\{Q^{2i},Q^{\dot2}_j\}&=4m\delta^i_j\cr
\{Q^i_i,Q^{2j}\}=\{Q^1_i,Q^{\dot2j}\}=0&(8.7)}$$
The first relation implies that
$$\langle
q^\mu_s|\big(Q^{1i}(Q^{1i})^*+(Q^{1i})^*Q^{1
i}\big)|q^\mu_s\rangle=0\eqno(8.8)$$ Demanding that the norm on
physical states be positive definite and
vanishes only if the state vanishes yields
$$Q_2^i|q^\mu_s\rangle=Q_{\dot2i}|q^\mu_s\rangle=0\eqno(8.9)$$
Hence, all generators in $H$ have zero action on rest-frame states
except $J,\ T_s,\ P_\mu,\ Q_1^i$ and $Q_{\dot1i}$. Using Eq. (2.23)
we  find that
$$\eqalignno{
[Q_1^i,J]&={1\over2}(\sigma_{12})_1^1Q_1^i\cr
&=-{i\over2}Q_1^i&(8.10)}$$
Similarly, we find that complex conjugation implies
$$[(Q_1^i)^*,J]=+{i\over2}(Q_1^i)^*\eqno(8.11)$$
The relations between the remaining generators summarised in Eqs.
(8.7), (8.10), (8.11) and (2.24) can be summarised by the statement
that $Q_1^i$ and $(Q_1^i)^*$ form a Clifford algebra, act as raising
and lowering operators for the helicity operator $J$ and transform
under the $N$ and $\bar N$ representation of $SU(N)$.

We find the representations of this algebra in the usual way; we
choose a state of given helicity, say $\lambda$, and let it be the
vacuum state for the operator $(Q_1^i)^*$, i.e.
$$\eqalignno{
Q_1^i|\lambda\rangle&=0\cr
J|\lambda\rangle&=i\lambda|\lambda\rangle&(8.12)}$$
The states of this representation are then
$$\eqalignno{
|\lambda\rangle&=|\lambda\rangle\cr
|\lambda-{1\over2},i\rangle&=(Q_1^i)^*|\lambda\rangle\cr
|\lambda-1,[ij]\rangle&=(Q_1^i)^*(Q_1^j)^*|\lambda\rangle&(8.13)}$$
etc. These states have the helicities indicated and belong to the
$[ijk\dots]$
anti-symmetric representation of $SU(N)$. The series will terminate
after the helicity $\lambda-(N/2)$, as the next state will be an
object  antisymmetric in $N+1$ indices. Since there
are only
$N$ labels this object vanishes identically. The states have
helicities from $\lambda$ to
$\lambda-(N/2)$, there being $N!/\big(m!(N-m)!\big)$ states with
helicity
$\lambda-(m/2)$.

To obtain a set of states which represent particles of both
helicities  we must add to the above set the representations with
helicities from
$-\lambda$ to $-\lambda+(N/2)$. The exception is the so-called CPT
self-conjugate sets of states which automatically contain both
helicity  states.

The representations of the full supersymmetry group are obtained by
boosting the above states in accordance with the Wigner method of
induced representations.

Hence the massless irreducible representation of $N=1$ supersymmetry
comprises only the two states
$$\eqalignno{&\quad|\lambda\rangle\cr
&|\lambda-{1\over2}\rangle=(Q_1)|\lambda\rangle&(8.14)}$$
with helicities $\lambda$ and $\lambda-{1\over2}$ and since
$$Q_1Q_1|\lambda\rangle=0\eqno(8.15)$$
there are no more states.

To obtain a CPT invariant theory we must add states of the opposite
helicities, i.e. $-\lambda$ and $-\lambda+{1\over2}$. For example, if
$\lambda={1\over2}$
we get on-shell helicity states 0 and ${1\over2}$ and their CPT
conjugates with helicities $-{1\over2},0$, giving a theory with two
spin 0's and one Majorana spin-${1\over2}$. Alternatively, if
$\lambda=2$ then we get on-shell helicity
states $3/2$ and 2 and their CPT self conjugates with helicity $-3/2$
and $-2$; this results in a theory with one spin 2 and one spin $3/2$
particles. These on-shell states are those of the Wess-Zumino model
and
$N=1$ supergravity respectively. Later in this discussion we will
give  a complete account of these theories.

For $N=4$ with $\lambda=1$ we get the massless states
$$|1\rangle,\ |{1\over2},i\rangle,\ |0,[ij]\rangle,\
|-{1\over2},[ijk]\rangle, \ |-1,[ijkl]\rangle\eqno(8.16)$$
This is a CPT self-conjugate theory with one spin, four
spin-${1\over2}$ and six spin-0 particles.

Table 8.1 below gives the multiplicity for massless irreducible
representations which have maximal helicity 1 or less.
\vfil\eject
\centerline{Table 8.1\quad Multiplicities for massless irreducible}
\centerline{representations with maximal helicity 1 or less}
$$\offinterlineskip\halign{\hfil$# $\hfil&\quad\strut#&\vrule\quad
\hfil$# $\hfil&\quad\hfil$# $\hfil&\quad\hfil$# $\hfil&\quad\hfil$#
$\hfil&\quad# \cr
\noalign{\hrule}
\qquad N&&&&&&\cr
{\rm Spin}&&1&1&2&2&4\cr
\noalign{\hrule}
{\rm Spin}\ 1&&-&1&1&-&1\cr
{\rm Spin}\ {1\over2}&&1&1&2&2&4\cr
{\rm Spin}\ 0&&2&-&2&4&6\cr
\noalign{\hrule}}$$
We see that as $N$ increases, the multiplicities of each spin and
the   number of different types of spin increases. The simplest
theories  are  those for $N=1$.
The one in the first column in the Wess-Zumino model and the one in
the second column is the $N=1$ supersymmetric Yang-Mills theory. The
latter contains one spin 1 and one spin ${1\over 2}$, consistent with
the formula for the lowest helicity $\lambda-(N/2)$, which in this
case  gives $1-{1\over 2}={1\over 2}$. The $N=4$ multiplet is CPT self
conjugate, since in this case we have $\lambda-(N/2)=1-4/2=-1$. The
Table stops at $N$ equal to 4 since when $N$ is greater than 4 we
must have particles of spin greater than 1. Clearly, $N>4$ implies
that $\lambda-(N/2)=1-(N/2)<-1$. This leads us to the well-known
statement that the $N=4$ supersymmetric theory is the maximally
extended Yang-Mills theory.

The content for massless on-shell representations with a maximum
helicity 2 is given in Table 8.2.
\par
\centerline{Table 8.2\quad Multiplicity for massless on-shell
representations with maximal helicity 2.}
$$\offinterlineskip\halign{\hfil$# $\hfil&\quad\strut#&\vrule\quad
\hfil$# $\hfil&\quad\hfil$# $\hfil&\quad\hfil$# $\hfil&\quad\hfil$#
$\hfil&
\quad\hfil$# $\hfil&\quad\hfil$# $\hfil&\quad\hfil$# $\hfil&\quad# \cr
\noalign{\hrule}
\qquad N&&&&&&&&&\cr
{\rm Spin}&&1&2&3&4&5&6&7&8\cr
\noalign{\hrule}
{\rm Spin}\ 2&&1&1&1&1&1&1&1&1\cr
{\rm Spin}\ {3\over2}&&1&2&3&4&5&6&8&8\cr
{\rm Spin}\ 1&&&1&3&6&10&16&28&28\cr
{\rm Spin}\ {1\over2}&&&&1&4&11&26&56&56\cr
{\rm Spin}\ 0&&&&&2&10&30&70&70\cr
\noalign{\hrule}}$$
The $N=1$ supergravity theory contains only one spin-2 graviton and
one spin-$ 3/2$ gravitino. It is often referred to as simple
supergravity theory. For the $N=8$ supergravity theory,
$\lambda-(N/2)=2-{8\over2}=-2$. Consequently it is CPT self conjugate
and contains all particles from spin 2 to spin 0. Clearly, for
theories  in which $N$ is greater than 8, particles of higher than
spin 2 will   occur. Thus, the $N=8$ theory is the maximally extended
supergravity    theory.

It has sometimes been  claimed that this theory is in fact the largest
possible   consistent supersymmetric  theory. This contention rests on
the widely-held belief   that it is impossible to consistently couple
massless particles of spin
${5\over2}$ to other particles.In fact
 superstring theories do include spin 5/2 particles, but these
are massive.

We now consider the massive irreducible representations of
supersymmetry. We take our rest-frame momentum to
be
$$q_s^\mu=(m,0,0,0)\eqno(8.17)$$
The corresponding little group is then generated by
$$P_m,Q^{\alpha
i},T^r,Z_1^{ij},Z_2^{ij},J_m\equiv{1\over2}\varepsilon_{mnr}J^
{nr}\eqno(8.18)$$
where $m,n,r=1,2,3$ for the present discussion. The $J_m$ generate
the  group $SU(2)$. Let us first consider the case where the central
charges  are trivially realised.

When acting on the rest-frame states the supercharges obey the
algebra
$$\eqalignno{\{Q^{Ai},(Q^{Bj})^*\}&=2\delta^A_B\delta^i_jm\cr
\{Q^{Ai},Q^{Bj}\}&=0&(8.19)}$$
The action of the $T^r$ is that of $U(N)$ with the $SU(2)$ rotation
generators satisfy
$$\eqalignno{[J_m,J_n]&=\varepsilon_{mnr}J_r\cr
[Q^{Ai},J_m]&=i(\sigma_m)^A_{\ B}Q^{Bi}&(8.20)}$$
where $(\sigma_m)$ are the Pauli matrices. We note that as far as
$SU(2)$ is
concerned the dotted spinor $Q^{\dot Ai}$ behaves like the undotted
spinor $Q_{Ai}$.

We observe that unlike the massless case none of the supercharges
are   trivially realised and so the Clifford algebra they form has
$4N$   elements, that is, twice as many as those for the massless
case. The   unique irreducible representation of the Clifford algebra
is found in   the usual way. We define a Clifford vacuum
$$Q_A^i|q^\mu_s\rangle=0,\quad A=1,2\ ,i=1,\dots , N
\eqno(8.21)$$
and the representation is carried by the states
$$|q^\mu_s\rangle,\ (Q_A^i)^*|q^\mu_s\rangle,\
(Q_A^i)^*(Q_B^j)^*|q^\mu_s\rangle,\dots\eqno(8.22)$$
Due to the anticommuting nature of the $(Q_A^j)^*$ this series
terminates when one applies $(2N+1)Q^*$'s.

The structure of the above representation is not particularly
apparent  since it is not clear how many particles of a given spin it
contains.  The properties of the Clifford algebra are more easily
displayed by  defining the real generators
$$\eqalignno{\Gamma^i_{2A-1}&={1\over2m}\big(Q^{Ai}+(Q^{Ai})^*\big)\cr
\Gamma^i_{2A}&={i\over2m}\big(Q^{Ai}-(Q^{Ai})^*\big)&(8.23)\cr
\noalign{where the}
\Gamma^i_p&=(\Gamma^i_1,\Gamma^i_2,\Gamma^i_3,\Gamma^i_4)&(8.24)}$$
are hermitian. The Clifford algebra of Eq. (8.19) now becomes
$$\{\Gamma_p^i,\Gamma_q^j\}=\delta^{ij}\delta_{pq}\eqno(8.25)$$
The $4N$ elements of the Clifford algebra carry the group $SO(4N)$ in
the standard manner; the $4N(4N-1)/2$ generators of $SO(4N)$ being
$$O^{ij}_{mn}={1\over2}[\Gamma_m^i,\Gamma_n^j]\eqno(8.26)$$
As there are an even number of elements in the basis of the Clifford
algebra, we may define a "parity" $(\gamma_5)$ operator
$$\Gamma_{4N+1}=\prod^4_{p=1}\prod^N_{i=1}\Gamma_p^i\eqno(8.27)$$
which obeys the relations
$$\eqalignno{
(\Gamma_{4N+1})^2&=+1\cr
\{\Gamma_{4N+1},\Gamma^i_p\}&=0&(8.28)}$$
Indeed, the irreducible representation of Eq. (8.22) is of dimension
$2^{2N}$ and transforms according to an irreducible representation
of
$SO(4N)$ of dimension $2^{2N-1}$ with $\Gamma_{4N+1}=-1$ and another
of  dimension $2^{2N-1}$ with $\Gamma_{4N+1}=+1$. Now any linear
transformation of the $Q$'s,
$Q^*$'s (for example $\delta Q=rQ$) can be represented by a
generator   formed from the commutator
 of the $Q$'s and $Q^*$'s (for example, $r[Q,Q^*])$. In particular
the
$SU(2)$ rotation generators are given by
$$s_k=-{i\over4m}(\sigma_k)^A_{\ B}[Q^{jB},(Q^{jA})^*]\eqno(8.29)$$
One may easily verify that
$$[Q^{jA},s_k]=i(\sigma_k)^A_BQ^{Bj}\eqno(8.30)$$
The states of a given spin will be classified by that subgroup of
$SO(4N)$ which commutes with the appropriate $SU(2)$ rotation
subgroup  of $SO(4N)$. This will be the group generated by all
generators  bilinear in $Q,Q^*$ that have their two-component index
contracted,  i.e.
$$\eqalignno{
\Lambda^i_j&={i\over2m}[Q^{Ai},(Q^{j}_A)^*]\cr
k^{ij}&={i\over2m}[Q^{Ai},Q_A^j]&
(8.31)}$$
and $(k^{ij})^\dagger=k_{ij}$. It is easy to verify that the
$\Lambda^i_j,\ k^
{ij}$ and $k_{ij}$ generate the group $USp(2N)$ and so the states of
a  given spin are labelled by representations of $USp(2N)$. That the
group  is $USp(2N)$
is most easily seen by defining
$$Q_A^a=\left\{\matrix{
Q_A^i\delta_i^a\qquad&a=1,\dots,N\qquad\cr
\varepsilon_{AB}(Q^{Bi})^*&a=N+1,\dots,2N\cr}\right.\eqno(8.32)$$
for then the generators $\Lambda^i_j,\ k^{ij}$ and $k_{ij}$ are given
by
$$s^{ab}={i\over2m}[Q^{Aa},Q_A^b].\eqno(8.33)$$
Using the fact that
$$\{Q_A^a,Q_B^b\}=\varepsilon_{AB}\Omega^{ab}\eqno(8.34)$$
where
$$\Omega^{ab}=\left(\matrix{
0&1\cr -1&0\cr}\right)$$
we can verify that
$$[s^{ab},s^{cd}]=\Omega^{ac}s^{bd}+\Omega^{ad}s^{bc}
+\Omega^{bc}s^{ad}+\Omega^{bd}s^{ac}\eqno(8.35)$$
which is the algebra of $USp(2N)$.

The particle content of a massive irreducible representation is
given   by the following

{\bf Theorem} [21]: If our Clifford vacuum is a scalar under the
$SU(2)$ spin group and the internal symmetry group, then the
irreducible massive representation of supersymmetry has the following
content
$$2^{2N}=\left[{N\over2},(0)\right]+\left[{N-1\over2},(1)\right]
+\dots+\left[{
N-\kappa\over2},(\kappa)\right]+\dots+[0,(N)]\eqno(8.36)$$
where the first entry in the bracket denotes the spin and the last
entry, say $(k)$ denotes which $k$th fold antisymmetric traceless
irreducible representation of $USp(2N)$ that this spin belongs to.

\centerline{Table 8.3 Some massive representations (without central
charges) labelled in}
\centerline{terms of the $USp(2N)$ representations.}
$$\offinterlineskip\halign{\hfil$# $\hfil&\quad\strut#&\vrule\quad
\hfil$# $\hfil&\quad\hfil$# $\hfil&\quad\hfil$# $\hfil&\quad\hfil$#
$\hfil&\quad\strut#&\vrule\quad\hfil$# $\hfil&\quad\hfil$#
$\hfil&\quad\hfil$# $\hfil
&\quad\strut#&\vrule\quad\hfil$# $\hfil&\quad\hfil$#
$\hfil&\quad\strut#&\vrule
\quad# \cr
\noalign{\hrule}
\qquad N&&&&&&&&&&&&&&\cr
{\rm Spin}&&&1&&&&&2&&&&3&&4\cr
\noalign{\hrule}
{\rm Spin}\ 2&&&&&1&&&&1&&&1&&1\cr
{\rm Spin}\ {3\over2}&&&&1&2&&&1&4&&1&6&&8\cr
{\rm Spin}\ 1&&&1&2&1&&1&4&5+1&&6&14+1&&27\cr
{\rm Spin}\ {1\over2}&&1&2&1&&&4&5+1&4&&14&14^\prime+6&&48\cr
{\rm Spin}\ 0&&2&1&&&&5&4&1&&14^\prime&14&&42\cr
\noalign{\hrule}}$$
\par
Consider an example with two supercharges. The classifying group is
$USp(4)$ and the $2^4$ states are one spin 1, four spin $1/2$, and
five  spin 0 corresponding to the ${\underline 1}-,\ {\underline 4}-$
and
${\underline 5}-$dimensional representations of $USp(4)$. For more
examples   see Table 8.3.
\par
Should the Clifford vacuum carry spin and belong to a non-trivial
representation of the internal group $U(N)$, then the irreducible
representation is found by taking the tensor product of the vacuum
and  the representation given in the above theorem.
\medskip

\centerline {\bf Massive Representations with a Central Charge}
\medskip
We now consider the case of particles that are massive, but which
also possess a central charge. We take the particles to be in their
rest frame with
momentum  $q^\mu\equiv (M,0,0,0)$. The
isotropy group, H  contains $(P^a, Q^{i}_A,Q_i^{\dot A},
\underline J, T_r$ and $ Z^{ij})$. In  the rest frame of the
particles, that is for the  momentum $q^\mu$, the algebra
of the  supercharges
is given by
$$\{Q^{A i},\ {(Q^{B j})}^* \}= 2\delta ^A_B \delta ^i_j M
\eqno(1)$$
and
$$\{Q_A^i,\ Q_B^j \}= \epsilon _{AB} Z^{ij}
\eqno(2)$$
\par
To discover what is the particle content in a supermultiplet
we would like to rewrite the above algebra as a Clifford
algebra. The first step in this proceedure is to carry out a unitary
transformation on the internal symmetry index of the supercharges
i.e. $Q_A^i\to U^i_j Q_A^j$ or $ Q_A\to U Q_A^i$ with $U^\dagger
U=1$. Such a transformation preserves the form of the first relation
of equation (1). However,  the unitary transformation can be chossen
[104] such that the central charge
,which transforms as $Z\to UZU^T$, can be brought to the form of a
matrix which has all its entries zero except for the 2 by 2 matices
down  its diagonal. These 2 by 2 matrices are anti-symmetric as a
consequence of the anti-symmetry nature
of $Z^{ij}$ which is preserved
by the unitary transformation. This is  the closest one can come to
diagonalising an anti-symmetric matrix. Let us for simplicity restrict
our attention to
$N$ even.  To best write
down this matrix we replace the
$i,j=1,2,\ldots N$ internal indices  by $i=(a,m), \ j=(b,n),\
a,b=1,2,\  m,n=1,\ldots ,{N\over 2}$ whereupon
$$Z^{(a,m) (b,n)} =
2 \epsilon ^{ab}\delta ^{mn}Z_n
\eqno(3)$$
In fact one also show that    $Z_n\ge 0$. The supercharges in the
rest frame satisfy the relations
$$ \{Q^{A (a m)},\ {(Q^{ B {(b n)}})}^*  \}= 2\delta ^A_B \delta
^a_b
\delta ^m_n M
\eqno(4)$$
and
$$\{Q_A^{(a m)},\ Q_B^{(b n) } \}= 2 \epsilon _{AB}
 \epsilon ^{ab}\delta ^{mn}Z_n
\eqno(5)$$
\par
We now define the supercharges
$$ S^{Am}_1 = {1\over \sqrt 2}(Q^{A1m}+ (Q^{B2m}\epsilon _{BA})^*)
\eqno(6)$$
$$
 S^{Am}_2 = {1\over \sqrt 2}(Q^{A1m}- (Q^{B2m}\epsilon _{BA})^*)
\eqno(7)$$
in terms of which all the anti-commutators vanish except for
$$
\{S^{Am}_1,\ {(S^{Bn}_1)}^*\}= 2 \delta ^{A B} \delta ^{mn}(M-Z_n)
\eqno(8)$$
$$
\{S^{Am}_2,\ {(S^{Bn}_2)}^*\}= 2 \delta ^{A B} \delta ^{mn}(M+Z_n)
\eqno(9)$$
This algebra is a  Clifford algebra
formed from the $2N$ operators $S^{Am}_1$ and
$S^{Am}_2$  and their complex conjugates. It follows from equation
(9) that if we take the same indices on each supercharge that
the right-hand side is positive definite and hence
$Z_n\le M $.
\par
To find the irreducible representation of supersymmetry we
follow a similar procedure to that which se followed  for  massive
and massless particles. The result crucially depends on whether
$Z_n<M,\ \forall\  n$ or if one or more values of n we
saturate the bound $Z_n=M$.
\par
Let us first consider  $Z_n<M,\ \forall\  n$. In this case,
the right-hand sides of both equations (8) and (9) are
non-zero. Taking
$S^{Am}_1$  and $S^{Am}_2$ to annihilate the vaccum
the physical states are given by the creation opperators
${(S^{Am}_1)}^*$  and
${(S^{Am}_2)}^*$ acting on the vacuum. The resulting representation
has
$2^{2N}$  states and has the same structure as for the massive case
in the absence of a central charge. The states are classified by
$USP(2N)$ as for the massive case.
\par
Let us now suppose  that  $q$ of the $Z_n$'s saturate the bound
i.e $Z_n=M$. For these values of $n$ the right-hand side of
equation (8) vanishes; taking the expectation value of this relation
for any physical state we find that
$$<phys| S^{An}_1 {(S^{An}_1)}^*|phys> +
<phys| {(S^{An}_1)}^*{(S^{An}_1)}|phys>=0
\eqno(10)$$
The scalar product on the space of physical states
satisfies all the axioms of a scalar product and hence we conclude
that both of the above terms vanish and as a result
$$ {(S^{Bn}_1)}^*|phys>=0={(S^{An}_1)}|phys>
\eqno(11)$$
This argument is the same as that used to eliminate
half of the supercharges and their complex conjugates in the massless
case, however in  case under consideration here
 it only eliminates
$q$ of the supercharges and their complex conjugates. There remain the
${N\over 2}$  supercharges ${(S^{Bm}_2)}$ and the ${N\over 2}-q $
supercharges
${(S^{Bm}_1)}$ for the values of $m$ for which $Z_m$ do not saturate
the bound as well as their complex conjugates. These supercharges
form a Clifford algebra and we can take the
${N\over 2}$  supercharges  ${(S^{Bm}_2)}$ and the
${N\over 2}-q $ supercharges ${(S^{Bm}_1)}$ to annihilate the vaccum
and their complex conjugates to be creation operators. The resulting
massive
irreducible representation of supersymmetry has $2^{2(N-q)}$
states and it has the same form as a massive representation
of $N-q$  extended supersymmetry. The states will be classified by
$USp(2N-2q)$.
\par
Clearly, a  representation in which some or all of
the central charges are equal to their mass has fewer states
 that the massive representation  formed  when none of the
central charges saturate the mass or a massive representation for
which  all
the central charges vanish. This is a consequence of the fact that the
latter  Clifford algebra has more of  its
supercharges active
in the  irreducible representation.
In almost all cases,
the representation with some of its central charges saturated
 contains  a smaller range of  spins than the massive  representation
with no central charges. This
feature plays a very important role in discussions  of duality in
supersymmetric theories.
\par
Let us consider the irreducible representations of   $N=4$
supersymmetry which has
both of its two  possible central charges saturated. These
representations are like the corresponding
$N=2$ massive representations. An important example
has  a ${\underline 1}$ of spin one, a ${\underline 4}$ of spin
one-half  and ${\underline 5}$ of spin zero.
The underlined numbers
are their
$USp(4)$ representations. This representation arises when
the $N=4$ Yang-Mills theory is spontaneously broken by
one of its scalars aquiring a vaccum expectation value. The
theory before being spontaneously broken has a  massless representation
with  one spin one, 2 spin 1/2 ,  and six spin zero's.
Examining the  massive representations for $N=4$ in the absence of a
central charge one finds that the  representation with the smallest
spins  has all spins from  spin 2 to  spin 0.   Hence the
spontaneously broken theory can only be supersymmetric if the
representation has a central charge.  Another way to get the count in
the above representation is to take the massless representation and
recall that  when the theory is spontaneously broken
one of the scalars
has been eaten by the vector as a result of the Higgs mechanism.

\par
We close this section by answering a question which may have arisen
in the mind of the reader. For the $N$ extended
supersymmetry algebra the supersymmetry algebra in the rest frame of
equation (3) representation has ${N\over 2 }$ possible central charges.
This makes  one central charge for the case of $N=2$. However, this
number conflicts with our understanding that a particle in $N=2$
supersymmetric Yang-Mills theory can have two central charges
corresponding to its electric and magnetic fields. The resolution of
this conundrum is that althought one can use a unitary  transformation
to bring the central charge of  a given irreducible  representation,
i.e. particle, to
have only one independent component one can not do this
simultaneously for all
irreducible multiplets or particles.
\par
Some examples of massive
representations with central charges are given
in the table below.
\medskip
\centerline{Table 8.4 Some massive representations with one central
charge $(|Z|=m)$.}
\centerline{All states are complex.}
$$\offinterlineskip\halign{\hfil$# $\hfil&\quad\strut#&\vrule\quad
\hfil$# $\hfil&\quad\hfil$# $\hfil&\quad\strut#&\vrule\quad\hfil$#
$\hfil&
\quad\hfil$# $\hfil&\quad\strut#&\vrule\quad\hfil$#
$\hfil&\quad\hfil$# $\hfil
&\quad\strut#&\vrule\quad# \cr
\noalign{\hrule}
\qquad N&&&&&&&&&&&\cr
{\rm Spin}&&2&&&4&&&6&&&8\cr
\noalign{\hrule}
{\rm Spin}\ 2&&&&&&&&&1&&1\cr
{\rm Spin}\ {3\over2}&&&&&&1&&1&6&&8\cr
{\rm Spin}\ 1&&&1&&1&4&&6&14+1&&27\cr
{\rm Spin}\ {1\over2}&&1&2&&4&5+1&&14&14^\prime+6&&48\cr
{\rm Spin}\ 0&&2&1&&5&4&&14^\prime&14&&42\cr
\noalign{\hrule}}$$

\par
 The account of
the  massive irreducible representations of supersymmetry given here
is  along similar lines to the review by Ferrara and Savoy given in
 [21].
\bigskip
{\centerline {\bf Lecture 4. Superspace }}
\bigskip
{\bf 4.1 Construction of Superspace}
\medskip
This constructed superspace as the coset space of the
super-Poincare group divided by the Lorentz group. It follows closely
chapter 14 of reference [0].
\bigskip
{\bf 4.2 Superspace Formulations of Rigid Supersymmetric Theories}
\medskip
The Wess-Zumino model and $N=1,2$ Yang-Mills theories are
formulated in superspace. This closely follows chapter 15 of reference
[0].
\bigskip
{\centerline {\bf Lecture 5. Quantum Properties of Supersymmetric
Models}}
\bigskip
{\bf 5.1 Super-Feynman Rules and the Non-renormalisation Theorem}
\medskip
The super-Feynman rules of the Wess-Zumino model and $N=1$
Yang-Mills theory are derived and the non-renormalisation theory is
proved. This closely followed chapter 17 of reference [0].
\bigskip
{\bf 5.2 Flat Directions}
\medskip
The potential in a supersymmetric theory  is given
by the  squares of the auxiliary fields. In this section we consider an
$N=1$ supersymmetric model which contains Wess-Zumino multiplets
coupled to the $N=1$ Yang-Mills multiplet with gauge group $G$.   Let
us  denote the auxiliary fields  of the Wess-Zumino multiplets  by
the complex field ${\cal F}^i$ where the index $i$ labels the
Wess-Zumino multiplets  and those of the $N=1$ Yang-Mills multiplet by
$ D^a$ where $a=1,\dots $, dimension of $G$. Then
the  classical potential is given by
$$
V= {|{\cal F}^i|}^2 +{1\over 2} \sum _a{( D^a)}^2
\eqno(5.2.1)
$$
For a general $N=1$ renormalizable theory the auxiliary fields are
given by
$$
{\cal F}^i= {\partial W (z^j)\over \partial z^i}
\eqno(5.2.2)
$$
and
$$ D^a= -g\bar z_i {(T^a )}^i_j z^j+ \zeta ^a
\eqno(5.2.3)
$$
In equation (5.2.2)  $W$ is the superpotential which we recall occurs
in the  superspace formulation of the theory as
$(\int d^4 x d^2 \theta W + c.c)$ and
$z^i$ are the scalars of the Wess-Zumino multiplet. For a renormlizable
theory, the  superpotential has
 the form $W(z)={1\over 3!} d_{ijk}z^i z^j z^k+  {1\over 2!} m_{ij}
z^i z^j+ e_iz^i$. In equation (5.2.3)  $g$ is the gauge coupling
constant and ${(T^a )}^i_j$
are the generators of the group $G$ to which these scalars
$z^i$ belong. The terms in the auxiliary fields which are
independent of $z^i$ can only occur when we have $U(1)$ factors for
$D^a$ and auxiliary fields ${\cal F}^i$ that transform trivial under
$G$. The resulting $\zeta^a$ and $e_i$ are constants.
\par
Clearly,  the potential is positive definite.
Another remarkable feature of the potential is that it
generically has flat  directions. This means  that
 minimizing  the potential  does not specify a unique  field
configuration. In other words  there exists a vacuum degeneracy. The
simplest example is for a Wess-Zumino model in the adjoint
representation coupled to a
$N=1$ Yang-Mills multiplet. Taking the superpotential for this theory
to vanish  the potential is given by
$$V= {1\over 2}\sum _a {(-g f_{abc}\bar z^b z^c)}^2
\eqno(5.2.4)
$$
Clearly, the minimum is given by field configurations whose only
non-zero vacuum expectation values are $ <z^a> H^a$ where
$H^a$ are the Cartan generators of the algebra. This theory is
precisely the
$N=2$  supersymmetric Yang-Mills theory when written in terms of $N=1$
supermuliplets.
\par
In a general quantum field theory such a vacuum degeneracy would be
removed by quantum corrections to the potential. However, things are
different in  supersymmetric theories. In fact,   if supersymmetry is
not broken  the potential does not
receive any perturbative quantum corrections [315]. It obviously
follows that     if supersymmetry is not broken
then the vacuum degeneracy is not removed by perturbative quantum
corrections [315]. This result was first proved
before the advent of the non-renormalisation theorem  as formulated
in  reference [316], but it is particularly obvious given this
theorem.  For the
effective potential we are interested in field configurations where the
spinors vanish and the space-time derivatives of all fields are set to
zero. For such configurations,   the gauge invariant
superfields do not contain any
$\theta$ dependence as only their first component is non-zero.
Quantum corrections, however,  contain an integral over all of
superspace and to be non-zero requires a $\theta ^2\bar \theta ^2$
factor in
the integrand.  For the field configurations of interest to us
such an integral over the full superspace must vanish and as a result
we find that there are no quantum corrections to the effective
potential if supersymmetry is not broken.
\par
Finally, we recall why the expectation values of the auxiliary
fields  vanish if supersymmetry is preserved. In this case
  the expectation
value of the supersymmetry  transformations  of the  spinors  must
vanish. The transformation of the spinors   contain auxiliary
fields which occur  without space-time
derivatives and the bosonic fields which correspond to the dynamical
degrees of freedom of the theory. The latter occur with space-time
derivative, as they have mass dimension one and $\epsilon$ has
dimension $-{1\over 2}$. Consequently, if the expectation
values of supersymmetry transformations of the spinors vanish so do the
expectation values of all the auxiliary fields. By examining the
supersymmetry transformations of the spinors given earlier the reader
may verify that there are no loop holes in this argument.
\par
Clearly, the rigid $N=2$  and $N=4$ theories can be written in terms
of $N=1$ supermultiplets and  so the flat directions that occur in
these theories are also  not removed by quantum corrections. Although
this might be viewed as a problem in these theories it has been
turned to advantage in the work of Seiberg and Witten. These authors
realized that   the  dependence of these theories on the expectation
values of the scalar  fields,   or the moduli,  obeyed
interesting properties  that can be exploited to solve for part of the
effective action of these theories.
\medskip
\bigskip
{\bf 5.3 Non-holomorphicity}
\medskip
The non-renormalisation theorem states that
perturbative quantum corrections to the effective action are of the
form
$$\int d^4x_1\ldots \int d^4x_n \int d^4\theta G(x_1, \ldots x_n)
f(\varphi(x_1,\theta), \ldots ,
V(x_1,\theta),\ldots ,D^A \varphi(x_1,\theta),\ldots )
\eqno(5.3.1)$$
where $\varphi$ and $V$ are the superfields that contain the
Wess-Zumino and $N=1$ Yang-Mills fields respectively.

The most significant aspect of this result is that the corrections
arise from  a  single superspace integral over {\bf all} of
superspace, that is they  contain a  $d^4\theta= d^2\theta d^2\bar
\theta $ integral and  not a  sub-integral
of the form $d^2\theta$ or $d^2\bar \theta$. Such sub-integrals play an
important role in supersymmetric theories. For example,
the superpotential in the superspace formulation of the Wess-Zumino
model has the  form $\int d^4 x d^2 \theta W + c.c$.
\par
While their is no question that this formulation of the
non-renormalisation theorem is correct,  with the passing of time, it
was taken by many workers to mean that their could never be any quantum
corrections which  were sub-integrals i.e. that is of the form
$$
\int d^4x  d^2\theta  \hat W(\varphi)
\eqno(5.3.2)
$$
In particular, it was often said that  there could be no quantum
corrections to the  superpotential.
\par
Consider, however, the expression
$$
\int d^4x  d^4\theta [{(-{D^2\over 4\partial ^2})}\varphi^n]
= \int d^4x  d^2\theta (-{\bar D^2\over 4})
 (-{D^2\over 4\partial^2})\varphi^n
 = \int d^4x d^2\theta \varphi^n
\eqno(5.3.3)
$$
where we have used the relation $\bar D^2 D^2\psi= 16 \partial ^2
\psi$ where $\psi$  is any chiral superfield. This maneuver
illustrates the important point that although  an expression
can be written  as a full superspace integral, it can also be
expressible as a local   integral over only a subspace of
superspace. The above expression when written in terms of the full
superspace integral is non-local, however, any  effective action
contains many  non-local contributions. The occurrence of the ${1\over
\partial^2}$ is the signal of a massless particle.  For  a massive
particle one would instead find  a factor of
${1\over (\partial^2+ m^2)}$ which can  not be rewritten as  a
sub-integral. Hence, only when massless particles circulate in the
quantum loops can we find a contribution to the effective action which
can be written as a  sub-superspace integral.
\par
The first example of such a correction to the superpotential was
found in reference [313].  In reference [301]. it was shown that all
the proofs of the non-renormalisation theorem allowed
contributions   to the effective action which were integrals over a
subspace of superspace if massless particles were present. It was also
shown [301] that  such corrections were not some pathological
exception, but that  they generically occurred whenever massless
particles were present. This lecture follows the first  part of
reference [301] and the reader is referred there for a much more
complete  discussion and several examples.  In the Wess-Zumino model
such corrections first occur at two loops and  were calculated in
[302], while in the Wess-Zumino model coupled to
$N=1$ Yang-Mills theory the corrections occur even at one loop
[303].  An alternative way of looking at such corrections was given in
references [304] and [305].
\par
The reader may wonder what such corrections have to do with
non-holomorphicity. The answer is that the corrections we have been
considering are non-holomorphic in the coupling constants. The
situation is most easily illustrated in the  context of the massless
Wess-Zumino model where  the superpotential is of the form
$\int d^4x  d^2\theta   \lambda \varphi^3 +c.c$. Since the propagator
connects $\varphi $ to
$\bar \varphi$
we get no corrections at all if we do not include terms that contain
both $\lambda$ and $\bar \lambda$.  Consequently, the
corrections we find  to the superpotential must  contain
$\lambda$ and $\bar \lambda$ and so is non-holomorphic in $\lambda$.
\par
We can of course prevent the occurrence of such terms if we give
masses to  all the particles or we  do not integrate
over the infra-red region of the loop momentum integration for the
massless  particles. Such is the case if we calculate the Wilsonian
effective action. However, if the terms considered here affect the
physics in an important way one will necessarily miss such effects and
they will only become apparent when one carries out the integrations
that one had previously excluded.
\bigskip
{\bf 5.4 Perturbative Quantum Properties of Extended Theories of
Supersymmetry}
\medskip
Many of the  perturbative properties of the extended theories of
supersymmetry are derived. These include the finiteness, or
superconformal invariance, of the
$N=4$ Yang-Mills theory,  the demonstration that $N=2$ Yang-Mills
theory coupled to $N=2$ matter has a perturbative beta-function
that only has one-loop contributions and  the existence of  a large
class of superconformally invariant  quantum $N=2$ theories.  This
section closely follows chapter 18 of reference [0].
\bigskip
{\bf {5.5 The $N=2$  Chiral Effective Action}}
\medskip
The $N=2$ Yang-Mills theory is described [58]
by a superfield $A$ which
is chiral
$$\bar D_{\dot B}^{\ i}{\cal A}=0
\eqno(5.5.1)$$
and also satisfies the constraint
$$D^{ij}{\cal A}=\bar D^{ij}\bar {\cal A}
\eqno(5.5.2)$$
where
$D^{ij}=D^{Ai}D_A^{\ j},\quad\bar D^{ij}=\bar D^{\dot Ai}\bar
D_{\dot A}^{\ j}$. This last constraint imposes the Bianchi identity
on the Yang-Mills field strength and makes the triplet auxiliary
field real.
\par
Let us decompose the $N=2$ chiral superfield ${\cal A}$  in terms of
$N=1$ superfields. Let us label the two superspace
Grassmann odd coordinates $\theta ^{A i},\ i=1,2$ which occur in the
$N=2$ superspace as
$\theta^A_{\ 1}=\theta^A$ and $\theta^A_{\ 2}=\eta^A$ and
$\bar \theta^{\dot A1}=\bar \theta^{\dot A}$ and $\bar \theta^{\dot
A2}=\eta^{\dot A}$. We associate the coordinates $\theta^A$ and
$\bar \theta^{\dot A}$ with those of $N=1$ superspace which we will
keep manifest.
Similarly, we denote the spinorial covariant derivatives of $N=2$
superspace as
$D_A^{\ 1}=D_A,\ D_A^{\ 2}=\nabla_A$ and $\bar D_{\dot A1}=
\bar D_{\dot
A},\ \bar D_{\dot A2}=\bar \nabla_{\dot A}$. We use also the notation
$$\bar D^2=\bar D^{\dot B}\bar D_{\dot B},\quad
D^2=D^BD_B,\quad\bar\nabla^2=\bar \nabla^{\dot B}\bar \nabla_{\dot
B},\quad\nabla^2=\nabla^B\nabla_B
\eqno(5.5.3)$$
We lower the $i,j$ indices with
$\varepsilon^{ij}=\varepsilon_{ji}=-\varepsilon^{ij},\
\varepsilon^{12}=1$, in the usual way, that is
$T^i=\varepsilon^{ij}T_j$ and $T_k=T^j\varepsilon_{jk}$.
\par
To find the $N=1$ superfields contained in ${\cal A}$ we must
write ${\cal A}$ in terms of $\eta_A, \eta_{\dot A}$  and
$N=1$ superfields and solve the
$N=2$ superspace constraints of equations (1) and (2).
The chirality constraint of equation (5.5.1) for  $i=2$
implies that ${\cal A}$ can be written in the form
$${\cal A}= a+\eta^A W_A+\eta^2K+\dots
\eqno(5.5.4)$$
where
$$\eta^2=\eta^A\eta_A\quad\bar\eta^2=\bar \eta^{\dot A}\bar \eta_{\dot
A}\quad\theta^2=\theta^A\theta_A\quad\bar\theta^2=\bar \theta^{\dot
A}\bar \theta_{\dot A}
\eqno(5.5.5)$$
and $+\dots$ denotes terms  involving  $\bar \eta^{\dot A}$ which
must contain space-time derivatives of the superfields that are
already written.  The superfields
${A},\ W_A$ and $K$ depend on
$x^\mu,\
\theta^A$ and
$\bar \theta^{\dot A}$ and so are $N=1$ superfields.
We will see that  $A$ and $W_C$ are the $N=1$ chiral
superfields that contain the Wess-Zumino multiplet and the
$N=1$ Yang-Mills mulitplet respectively.
\par
We could continue with this approach, however a more efficient
method  is to use the covariant derivatives to define the
components of the superfield ${\cal A}$.
Acting  with $\nabla_B$ on
${\cal A}$, equation (5.5.1) for $i=2$ implies  that the only
independent
$N=1$ superfields which depend on $x^\mu,\theta _A$
and
$\bar \theta_{\dot A}$ are
are given by
$${\cal A}|_{\eta=0}=A,\quad
\nabla_B A|_{\eta=0}=W_B,\quad-{1\over2}\nabla^2A|_{\eta=0}=K.
\eqno(5.5.6)$$
 Acting with $\bar D_{\dot C}$, equation (5.5.1)
for $i=1$
implies that these  $N=1$ superfields   are
 chiral superfields;
$$\bar D_{\dot B}A=0=\bar D_{\dot B}W_C=\bar D_{\dot B}K
\eqno(5.5.7)$$
\par
It remains to solve the constraint of equation (5.5.2).
Taking $i=1,\ j=2$
we find it becomes
$D^B\nabla_BA=-\nabla^{\dot B}D_{\dot B}A$.
Swopping the last two covariant derivatives on the right hand side and
taking $\eta=0$ we find the constraint
$D_B W^B=\bar D_{\dot B} \bar W^{\dot B}$.
Taking $i=2,\ j=2$ in equation (5.5.2) we find the result
$$\nabla^2A=\bar D^2\bar A
\eqno(5.5.8)$$
Evaluating this equation at $\eta=0$ we find that
$$-{1\over2}\nabla^2 A|_{\eta=0}=K=-{1\over2}\bar D^2\bar A
\eqno(5.5.9)$$
Hence $K$ is not an independent superfield.
\par
To summarise; the $N=2$ superfield ${\cal A}$ decomposes  into
two $N=1$ superfields
 $A$ and $W_B$ which are  subject to the
constraints
$$\bar D_{\dot B} A=0,\ \bar D_{\dot B}W_c=0
,\ D^BW_B=\bar D^{\dot B}\bar W_{\dot B}
\eqno(5.5.10)$$
These $N=1$ superfields contain the Wess-Zumino  and $N=1$
Yang-Mills  multiplets.
\par
We can further define the $x$-space component
superfields as follows
$$A|_{\theta=0}= a,\ D_B A|_{\theta=0}=
\chi_B,\  -{1\over 2}D^2A|_{\theta=0}= f
\eqno(5.5.11)$$
and
$$W_B|_{\theta=0}= \lambda_B,\ D_CW_B|_{\theta=0}=
{ {\cal F}^{\mu\nu}} (\sigma _{\mu\nu})_{BC}+iD\epsilon_{CB}
\eqno(5.5.12)$$
where ${ {\cal F}^{\mu\nu}}= F^{\mu\nu}-i{}^*F^{\mu\nu}$ and
${}^*F^{\mu\nu}= {1\over 2} \epsilon _{\mu\nu\rho\lambda}
F^{\rho\lambda}$
\par
Let us now consider an $N=2$ effective theory whose only massless
particles are $N=2$ $U(1)$ multiplets. In such a theory we can
integrate over all the massive particles in
the functional integral to find an effective action for the remaining
$U(1)$ supermultiplets.  The most simple example is   the
$SU(2)$ $N=2$ Yang-Mills theory that is spontaneously broken to
$U(1)$ by shifting one of the scalar fields of the theory. Although
the usual action for such
$U(1)$ mulitplets is a free theory, the effective action resulting
from such a process  is very non-trivial and must be of the form
$$ Im\int d^4xd^4\theta F({\cal A}) +\int d^4 xd^4\theta d^4\bar \theta
K({\cal A},\bar {\cal A}, D_B{\cal A},\ldots )
\eqno(5.5.13)$$
For simplicity we have suppressed the index which would label the
different $U(1)$ factors.
\par
It can be argued as follows that the low energy part of
such an action is just given by the first term. The integrand of the
second term has mass  dimension zero. As such, one must
introduce a mass scale
$\Lambda$ and it will contain terms such as  ${{\cal A}^{2n}\over
\Lambda^{2n}},n\in\ {\bf Z}$. Evaluating this  in terms of the
$x$-space component fields we find terms of the  form
$$\int d^4 x {a^n(\partial_\nu
\partial ^\nu)^{4}\bar a^n\over \Lambda ^{2n}}+\dots
\eqno(5.5.14)$$
where $+\dots$ means
its supersymmetric completion. Although one can consider more
complicated contributions to  the  effective action, the final result
must seemingly contain the  mass scale
$\Lambda $ to the appropriate
power and so be higher order in derivatives than the kinetic energy
term for the fields. This is to be contrasted with  the
first term of equation (5.5.13), the integrand of which  has mass
dimension  two and so the resulting
$x$-space expressions are of the same order in derivatives as the
standard kinetic energy terms.  Thus if we are only interested in the
low energy effective theory one can neglect the  second term
and only consider the  chiral effective action is of the form
$$Im\int d^4xd^4\theta F({\cal A})
\eqno(5.5.15)$$
\par
The above argument as given applies not just to $U(1)$
factors, but to any $N=2$ theory. In fact, there is a flaw in this
argument; the integrand of the second term of equation (5.5.11) could
contain terms of the form
$ln ({{\cal A}_1{\cal A}_2\over {\cal A}_3 {\cal A}_4})$ where ${\cal
A}_i,\ i=1,2,3,4$ are different  superfields for the $U(1)$ factors
or other superfields in the theory.   Clearly, this term does not
involve a mass scale
$\Lambda$  and its
$x$-space expression will not be higher order in derivatives.  Indeed
one finds [321],[322] that  just  such
terms arise even  in the one loop calculation of
$N=2$ $SU(2)$
Yang-Mills theory before it is spontaneously broken. Fortunately they
can not occur in the effective action for the  $N=2$ $U(1)$
theory resulting from the spontaneous breaking of the $N=2$ $SU(2)$
Yang-Mills theory discussed above.
\par
Let us now assume that we are only interested in a low energy
effective theory whose action is of the form given in equation
(5.5.15).
The solution to the above constraints of equation (5.5.1) and (5.5.2)
for the Abelian theory are of the  form [64]
$${\cal A}=\bar D^4D^{ij}V_{ij}
\eqno(5.5.16)$$
where $V_{ij}$ is an unconstrained superfield of mass dimension -2.
Varying the above action with respect to $V_{ij}$ yields the equation
of motion. We find that
$$Im \int d^4xd^4\theta\bar D^4D^{ij}\delta V_{ij}{dF({\cal A})\over
d{\cal A}}\propto Im \int d^4xd^8\theta\delta V_{ij}D^{ij}{dF\over
d{\cal A}}
\eqno(5.5.17)$$
and hence we find the equation of motion to be
$$D^{ij}{dF\over d{\cal A}}=\bar D^{ij}{d\bar F\over d\bar {\cal A}}
\eqno(5.5.18$$
We observe that if we write ${\cal A}_D={dF\over d{\cal A}}$ and ${\cal A}$
as a doublet
i.e. $\left(\matrix{{\cal A}_D\cr {\cal A}\cr}\right)$. Then
the equation (5.5.2) which
encodes the Bianchi identity and the equation of motion (5.5.18) can be
written as
$$D^{ij}\left(\matrix{{\cal A}_D\cr {\cal A}\cr}\right)=\bar
D^{ij}\left(\matrix{{\cal A}_D\cr {\cal A}\cr}\right)
\eqno(5.5.19)$$
Clearly, this system has a set of equations of motion invariant under
$$\left(\matrix{{\cal A}_D\cr
{\cal A}\cr}\right)
\rightarrow\Omega\left(\matrix{{\cal A}_D\cr {\cal A}\cr}\right)
\eqno(5.5.20)$$
where $\Omega\in SL(2,R)$. This symmetry is restricted to
 $SL(2,{\bf Z})$ when the theory is quantized. It is clear form the
free theory that this symmetry interchanges the equation of motion with
the Bianchi identity and thus is a duality trasnformation.
This symmetry was first discussed in this theory in reference [324].
The $N=2$ superspace formulation  of the theory makes the
appearance of this symmetry particularly apparent [325].
\par
For the free theory $F=i{\cal A}^2$ and we find the
equation of motion and
constraint imply that $D^{ij}{\cal A}=0=\bar D^{ij}{\cal A}$.
As a result at $\theta^{{\cal A}i}=0$ the
triplet auxiliary fields of the $N=2$ Yang-Mills theory  vanish
ensuring the correct equations of motion.
\par
We can now evaluate the chiral effective action [318]:
$$\eqalign{
Im\int  d^4xd^4\theta F&=Im \int d^4xd^2\theta d^2\eta F=\int
d^4xd^2\theta\left(-{\nabla\over4}\right)^2F\cr
&=Im \int d^4xd^2\theta\left\{-{(\nabla^2{\cal A})\over4}{dF\over
d{\cal A}}-{1\over 4}\nabla^{ B}{\cal A}\nabla_{ B}{\cal
A}{d^2F\over d{\cal A}^2}\right\}_{\eta=0}\cr
&=Im \int
d^4xd^2\theta\left\{(-{1\over 4}\bar D^2\bar A){dF\over
d A}
-{1\over 4}{d^2F\over d{ A}^2}W^{ B}W_{ B}\right\}\cr &=
Im \int d^4xd^4\theta\bar A{dF\over d A}
-{1\over 4}\int
d^4xd^2\theta{d^2F\over d{ A}^2}W^{B}W_{B}}
\eqno(5.5.21)$$
One can further evaluate the above action in terms of the
$x$-space component fields by writing  the $d^2 \theta
= -{1\over 4}D^2$ and $d^2 \bar \theta
= -{1\over 4}\bar D^2$ and using the definitions
of the $x$-space component fields of equations (5.5.11) and (5.5.12).
The result is
$$ A_1+A_2
\eqno(5.5.22)$$
where the fermion independent part is given by
$$ A_1= Im \int d^4 x {d^2 F\over d a^2}\{- \partial _\mu \bar a
\partial ^\mu a   -  {\cal F}_{\mu\nu}
{\cal F}^{\mu\nu}+ {1\over 4}(f\bar f + D^2)
\}
\eqno(5.5.23)$$
and
$$ A_2= Im \int d^4 x\{
-{i\over 4}{d^2 F\over d a^2}(\bar \chi _{\dot
A }(\sigma ^\mu)^{B\dot A}\partial_\mu \chi _B +
\lambda  _{\dot A}
(\sigma ^\mu)^{B\dot A}\partial_\mu \lambda _B)$$
$$
-  {1\over 8}{d^3 F\over d a^3}
(\bar f\chi^A \chi_a + f\lambda ^A \lambda _A
-iD\chi^A\lambda_A)
-{1\over 8}{d^3 F\over d a^3}\chi^A\lambda^B (\sigma _{\mu\nu}
)_{AB}{\cal F}_{\mu\nu}
+ {1\over 16}{d^4 F\over d a^4} \chi^A\chi_A\lambda ^A \lambda _A)
\eqno(5.5.24)$$
\par
We now wish to find the complete form of the
perturbative expression for the chiral effective action of equation
(5.5.15). This result can be derived  as a consequence of the way the
theory breaks superconformal invariance. For simplicity let us begin
with the theory of an $N=2$ $U(1)$ mulitplet which is a free theory.
In this case, the effective action and the original action coincide
and are given by
$F=i{{\cal A}}^2$. This action  is invariant
under the $N=2$ superconformal group discussed in   section 6. In
particular, it is invariant under the internal
$U(2)=SU(2)\otimes U(1)$ transformations. The $U(1)$ transformations
are called $R$ transformations and they can be taken to  act on the
Grassmann odd coordinates $\theta $ as
$\theta \to  e^{i\alpha}\theta $ where $\alpha $ is the
parameter of the transformation (see section 6). The $R$
transformations  act on
${\cal A}$ as ${\cal A}\to e^{2i\alpha}{\cal A}$. It is easy to see
that with the choice of $F=i{{\cal A}}^2$ then $F\to e^{4i\alpha} F$
which cancels the variation of $\int d^4\theta$ leaving the
action invariant. It is straightforward to deduce the action
of $R$ transformations on the component fields, it acts as a chiral
rotations on the spinors $\lambda$ and  acts on the scalars $a$
as $a \to e^{2i\alpha} a$. The  dilations can be handled in a similar
manner. Under a dilation with parameter $w$
one finds that $x^\mu \to e^{-w} x^\mu, \theta ^A \to e^{{w\over
2}}\theta ^A$ and ${\cal A}\to e^{2w}{\cal A}$ which leaves invariant
the free action.
\par
The $N=2$ superconformal currents are contained in the superfield $J$
which is subject to the constraint
$$ D^{ij} J=0
\eqno(5.5.25)$$
For the free $U(1)$ theory discussed just above
$J= {\cal A}\bar {\cal A}$ and we find that it does indeed satisfy
the above equation  as $D^{ij} ({\cal A}\bar {\cal A})= D^{ij} ({\cal
A})\bar {\cal A}=0$ by virtue of the equation of motion of ${\cal A}$.
The superfield $J$ has mass dimension two, $R$ weight $0$ and the
$U(2)$ Currents are the $\theta^A=0$ component of $D^i_A\bar D^j_{\dot
A} J$.
\par
For the  theories we are considering  the effective
action does not in general possess a superconformal symmetry as the
underlying  theory is anomalous. The anomaly must modify equation
(5.5.25) by a term  on the right hand side that is constructed from
${\cal A}$ and has mass dimension three, $R$ weight $-2$ and
the correct $SU(2)$ transformation property. The only possible choice
is [121]
$$  D^{ij} J= -{1\over 3}{\beta(g)\over g}\bar D^{ij}({\bar {\cal
A}})^2
\eqno(5.5.26)$$
where $\beta(g)$ is a function of the gauge coupling constant $g$
that will turn out to be the $\beta$-function, but at this point is
an unknown quantity.  At first sight, one might think one could
add a term of the form $ \bar D^{ij}({\bar {\cal A}}){\cal A}$ however
using the Bianchi identity of equation (5.5.2) we can manipulate this
expression  to be of the form of $J$ and so absorb it on the left
hand side  as a redefinition of the current. A review of
supercurrent multiplets and their anomalies is given in
chapter 20 of reference [0].
\par
By acting with spinorial covariant derivatives on $J$ in equation
(5.5.26) we find that it implies that  the divergence of the
$R$ current $j_\mu^5$ is of  the form
$$\partial ^\mu j_\mu^5= {2\over 3}{\beta(g)\over g}(-{1\over 4}
F^{\mu\nu}{}^* F_{\mu\nu}+\dots)
\eqno(5.5.27)$$
and corresponding to the breaking of dilations the trace of the
energy-momentum tensor
$$\theta ^\mu_\mu= -2{\beta(g)\over g} (-{1\over 4} F^{\mu\nu}
F_{\mu\nu}+\dots)
\eqno(5.5.28)$$
In the these equation $+\dots $ means terms involving fermions and
scalars. It
is known [320] that the latter equation for $\theta ^\mu_\mu$ is
proportional to  the beta-function  times
$F^{\mu\nu} F_{\mu\nu}$. Taking into account the constant of
proportionality we conclude that $\beta$ is indeed that
the  beta-function of the theory.
\par
It was shown in references [120] with related work in
references [121],[114],[115] that  the perturbative contribution to
beta function in any renormalizable
$N=2$ Yang-Mills theory  coupled to $N=2$ matter
 has only  a one loop contribution. The derivation of this
result is too complicated to give here, but we refer the reader
to chapter 18 of  reference [0] for a review. The full perturbative
beta function in such a theory is then  [120]
$$\beta ={g^3\over (4\pi)^2}(T(R_\sigma)-3C_2(G))\equiv
{3 g^3\over 4 \pi^2}\beta _1
\eqno(5.5.30)$$
Here $(T^s)^a_{\ b}(T^s)^b_{\ a} =\delta ^{st}T(R)$ where
$(T^s)^a_{\ b}$
are the generators of the gauge group $G$ in the representation $R$
to which the $N=2$ matter belongs and $f_{rst}f_{qst}=\delta_{rq}
C_2(G)$ where $f_{rst}$ are the structure constants of the group $G$.
For example for the case of $SU(N)$ we find that $C_2=N$ and that
if the matter is in the fundamental representation $T(R)={1\over 2}$.
The reader may readily construct some of the finite quantum field
theories discovered in reference [120].
\par
Finally we can consider the
constraints placed on the perturbative contribution to the effective
action by  the anomaly equation (5.5.26). In particular, let us
consider the  underlying theory to be an $SU(2)$ $N=2$
 Yang-Mills theory  coupled to $N=2$ matter
which is spontaneously broken to $U(1)$. The beta
function that occurs in equation (5.5.26) is that for the underlying
theory and  so in these  equations  we can substitute
the beta function of  equation (5.5.30).
  The integrated Ward identity
tells us that the variation of the  effective action is equal to the
anomaly. Carrying  this out for
the dilations and
$R$ transformations one finds that
$$\delta (Im \int  d^4xd^4\theta F)= Im \int  d^4xd^4\theta
(2i\alpha +w)( {\cal A} {d F\over d {\cal A}}- 2F)= Im
\int  d^4xd^4\theta
(2i\alpha+w) 8\pi i \beta _1 u
\eqno(5.5.31)$$
where $u=-{A^2\over 4\pi^2}$. Although this equation is
very natural in
that both sides have  chiral integrands and are of the correct
dimension and $R$ weight it may be unclear to the reader that the
right-hand side of  this equation  should really be of this precise
form. To derive the equation one must formulate the
Ward-identity for the $N=2 $ theory. The Ward identity contains three
terms the variation of the effective action, the divergence condition
for the currents and the anomaly term. To find the above integrated
form  we must  integrate the Ward identity  in such a
 way  as to eliminate  the term which depends on the
supercurrent. This process is rather complicated, however the reader
may verify that  at the component level the above equation  does
contain the right hand side of equations (5.5.27) and (5.5.28) in the
  way dictated from the  usual Ward identity in $x$-space.  Only by
carrying out this proceedure can one determine the somewhat elisve
constant. From  equation (5.5.31) we can extract the equation
$$  {\cal A} {d F\over d {\cal A}}- 2F =  8\pi i \beta_1 u
\eqno(5.5.32)$$
 Upon  making the substitution
$F={\cal A}^2 G$ and solving the resulting equation for $G$
 we find that [319]
$$F=    -{i\over \pi}\beta _1{\cal A}^2ln {{\cal A}^2\over \Lambda^2}
\eqno(5.5.33)$$
\par
Thus using  the anomaly equation and the knowledge
of the perturbative beta function
in the $N=2$  theories [120]
we can determine the perturbative  part
of the chiral effective action completely [319].
\par
In the above we have used the fact that in a supersymmetric theory
the currents of symmetries of the theory belong in a supermultiplet.
As such,  knowledge  of the properties about one of the currents can
be used to deduce the related behaviour of the other currents and
so deduce consequences for
the theory. This type of argument was first used [112]
to show that the $N=4$ Yang-Mills theory is finite [112-117].
\par
In fact, equation (5.5.32) holds for the full non-perturbative
quantum theory except in this case the anomaly (i.e. $u$) is not
  the simple function of ${\cal A}$ given above. The
demonstration  of this fact and a more careful derivation of the last
equations given in this section can be found in reference [323].
\bigskip
{\centerline {\bf Lecture 6. Superconformal Theories}}
\bigskip
{ \bf 6.1 The Geometry of Superconformal Transformations}
\medskip
The  superspace that we used in lecture 4 was defined as the coset
space of the super-Poincare group  divided by
the Lorentz group and internal symmetry group. This
superspace is called Minkowski superspace.  For the case of the
$N=1$ super-Poincare group, the  superspace is  parameterised by the
coordinates
$Z^M=( x^\mu, \theta ^A, \theta ^{\dot A})$ corresponding to the
generators $P^\mu$ and $Q ^A, Q ^{\dot A}$ which generate
transformations that are not contained in the isotropy subgroup. We
can construct on superspace  a set of preferred frames with
supervierbeins
$E_M^{\phantom {M}\pi}$ . The covariant derivatives are given by
$D_M= E_M^{\ \pi}{\partial \over \partial Z^\pi}$. Their precise form
being
$$D_A= {\partial \over \partial \theta ^A}-i(\sigma ^m)_{A{\dot B}}
\theta^{\dot B}\partial_m,\
 D_{\dot A}= {\partial \over \partial \theta ^{\dot A}}-i(\sigma
^m)_{B{\dot  A}}
\theta^{B}\partial_m, \eqno(6.1)$$
and
$$D_m= {\partial \over \partial x^m}\equiv \partial _m
\eqno(6.2)$$
We can read off the components of the inverse supervierbien from
these equations.
\par
For  superconformal theories it is more
natural to consider  a superspace which is constructed from  the
coset space found by dividing the  superconformal group by  the
subgroup which is generated by Lorentz transformations $J_{\mu\nu}$,
dilations $D$, special translations $K_\mu$ and special supersymmetry
transformations $S_{A i}, S_{\dot A i}, i=1,\dots ,N$ and the internal
symmetry generators. The internal group for the  superconformal
algebra
 contains the group $U(N)= SU(N) \times U(1)$
although in the case of $N=4$ the $U(1)$ factor does not act on the
supercharges.

This coset construction   leads to the same Minkowski superspace
with the   same transformations for the super-Poincare group, but it
has the advantage that it automatically encodes the action of   the
superconformal transformations on the superspace.   These
transformations were first calculated by Martin Sohnius in reference
[306].
\par
The purpose of this section is  to give an alternative method of
calculating the superconformal transformations in four
dimensions which   will enable us to give a compact superspace
form for the superconformal transformations. In particular,
 all the parameters of the transformations will be  encode in one
superfield which we can think of as the superspace equivalent of a
conformal Killing vector. This  formulation was first given by
B. Conlong and P. West and can be found in
reference [307]. One reason for reviewing  this work  here is that
there  is still not  a readable account readily available in the
literature. This section was written in collaboration with B.
Conlong.  Some    reviews on this subject can be found in [308].

 Conformal transformations in Minkowski space are defined to be those
transformations which preserve the Minkowski metric up to scale (
see chapter 25  of reference [0] for a review). However, superspace
does not have a natural metric since the tangent space group contains
the Lorentz group which does not relate the bosonic to the
fermionic sectors of the tangent space.  The treatment we now give
follows that given in chapter 25 of reference [0] for the case of two
dimensional superconformal transformations.
\par
There are two methods to define a  superconformal transformation.
\par
\item {[a]}
 We can demand that it is a superdiffeomorphism
which preserves part of the bosonic part of the supersymmetric line
element
$$ dL^m d L^n \eta _{mn}
\eqno(6.3)$$
where
$$dL^m\equiv  dZ^M E_M^{\ m} = dx^m +i\bar \theta \gamma^m d \theta
\eqno(6.4)
$$
 up to an arbitrary local scale factor.
\par
\item {[b]}
 We can alternatively  demand that it is a superdiffeomorphism
which preserves  the  spinor components of the
superspace covariant derivatives up to an arbitrary local scale
factor.  More precisely,  a superconformal
transformation is one such that
$$
D_A\mapsto D'_A=f_A^{\phantom{A}B}(z)D_B,\ \
\bar D_{\dot A}\mapsto\bar D'_{\dot A}=\bar f_{\dot
A}^{\phantom{A}\dot B}(z)\bar D_{\dot B}
\eqno(6.5)$$
We note that the transformation must preserve each chirality
spinor derivative separately.
In fact, these two definitions are equivalent and we
will work  with only the second definition.

Carrying out the super-reparameterisation
$z^M=(x^\mu,\theta^A,\bar\theta^{\dot
A})\mapsto z^{M\prime }=(x^{\mu \prime },\theta^{A \prime},
\bar\theta^{\dot A \prime })$
upon the spinorial covariant derivatives we find that a
finite superconformal transformation obeys the constraints
$$
D_A\bar\theta'^{\dot B}=0,\ \
D_Ax'^\mu+i\sigma^\mu_{B\dot B}\bar\theta'^{\dot B}(D_A\theta'^B)
=0
\eqno(6.6)
$$
and
$$\bar D_{\dot A}\theta'^B=0,\ \
\bar D_{\dot A}x'^\mu+i\sigma^\mu_{B\dot B}\theta'^B(\bar D_{\dot A}
\bar\theta'^{\dot B})=0
\eqno(6.7)$$
The corresponding transformation of the covariant derivatives being
$$
D_A=(D_A\theta'^B)D'_B,\ \
 \bar D_{\dot A}=(\bar D_{\dot A}\bar\theta'^{\dot B})\bar D'_{\dot
B}
\eqno(6.8)$$
\par
We now  consider  an infinitesimal transformation
$$z^\pi\mapsto z'^\pi=z^\pi+G^\pi
\eqno(6.9)$$
where $G^\pi=(G^\mu,G^A,G^{\dot A})$  is a set of infinitesimal
superfields. Equations (6.6) and (6.7) then  become
$$
D_A G^{\dot A}=0,\ \
D_A G^\mu+i\sigma^\mu_{A\dot B}G^{\dot B}+
i\sigma^\mu_{B\dot B}\bar\theta^{\dot B}D_A G^B=0
\eqno(6.10)$$
and
$$\bar D_{\dot A}\bar G^B=0,\ \
\bar D_{\dot A}G^\mu+i\sigma^\mu_{A\dot A}G^A+i\sigma^\mu_{B\dot B}
\theta^B\bar D_{\dot A}\bar G^{\dot B}=0
\eqno(6.11)$$
\par
The vector field corresponding to such an infinitesimal
transformation is given by  $V= G^\pi\partial _\pi$. However, this  can
also be written as   $V= F^M D_M$
where the change of basis corresponds to the relation $F^M= G^\pi
E_\pi^M$. In terms of components this change is given by
$$
 F^\mu= G^\mu-i\sigma^\mu_{A\dot A}\bar\theta^{\dot A}
G^A-i\sigma^\mu_{A\dot A}\theta^A G^{\dot A}
\eqno(6.12)$$
as well as
$$F^A= G^A,\ F^{\dot A}= G^{\dot A}
\eqno(6.13)$$
We shall denote the vector component by $F^\mu,\ldots$ or $F^n,\ldots$
  even though it should strictly speaking carry the latter
$m,n,\ldots $ indices.  It is straightforward to verify that
equations (6.10) and (6.11) now take the  neater form
$$
D_A  F^\mu  = -2i\sigma^\mu_{A\dot A}\bar F^{\dot A},\
\bar D_{\dot A} F^\mu  = -2i\sigma^\mu_{A\dot A} F^A
\eqno(6.14)
$$
and
$$D_{\dot A}F^B=0= D_AF^{\dot B}.
\eqno(6.15)$$
\par
A somewhat quicker derivation of this result can be given by first
writing the infinitesimal change in the covariant derivatives
 under an infinitesimal superdiffeomorphism  in the form
$D_M\to D_M+ [V,D_M]$.  Using  the form for $V$ given above which
contains the covariant derivatives and then using  the fact that the
only non-zero commutator or anti-commutator, where appropriate, of
the covariant derivatives is $\{ D_A,D_{\dot B}\} = -2i
{(\sigma^n)}_{A\dot B}\partial_ n$ we recover equations (6.14) and
(6.15).
\par
Equation (6.14) can be rewritten as
$$
F^{\dot A}=-{1\over 8i}\sigma_\mu^{A\dot A}D_A  F^\mu,\
F^A=-{1\over 8i}\sigma_\mu^{A\dot A}\bar D_{\dot A}F^\mu
\eqno(6.16)$$
from which it is apparent that  all transformations may be
expressed in terms of
$ F^\mu$ alone. Using equations (6.12) and (6.13) we find that the
explicit transformations of the coordinates are given by
$$\eqalign{
x'^\mu&=x^\mu+( F^\mu-{1\over 8}\theta^A D_A
F^\mu+{1\over
8}\theta^A{(\sigma^\mu_{\ \nu})}_{AC} D^C
F^\nu-{1\over 8}\bar\theta^{\dot A}\bar D_{\dot A} F^\mu+{1\over
8}\bar\theta^{\dot A}{(\bar\sigma^\mu_{\ \nu})}_{\dot A\dot
C}\bar D^{\dot C} F^\nu)\cr
\theta'^A&=\theta^A-{1\over 8i}\sigma_{\mu}^{A\dot A}\bar D_{\dot
A} F^\mu\cr
\bar\theta'^{\dot A}&=\bar\theta^{\dot A}-{1\over 8i}
\sigma_{\mu}^{A\dot A}D_A
 F^\mu \cr }\eqno(6.18)$$
 Let us  define $F_{B\dot B}\equiv \sigma^\mu_{B\dot B} F_\mu$,
whereupon  equation (6.14) becomes
$$
D_A F_{B\dot B}=-4i\varepsilon_{AB}F_{\dot B},\ \
\bar D_{\dot A} F_{B\dot B}=-4i\varepsilon_{\dot A\dot B}F_B
\eqno(6.19)$$
from which we may deduce the constraint
$$D_{(A}F_{{B)\dot B}}=0,\ \ \bar D_{(\dot A}F_{\vert B\vert\dot B)}=0
\eqno(6.20)$$
Acting with $D_C$ on $D_AF_{B\dot B}$ and using equation (6.19) we
conclude that
$$D_CD_AF_{B\dot B}= -4i\varepsilon_{ A B}D_CF_{\dot B}=
- D_AD_CF_{B\dot B}= -4i\varepsilon_{ C B}D_AF_{\dot B}=0.
\eqno(6.21) $$
The last step follows by tracing with  $\varepsilon^{ A B}$.
 Consequently, we find that equation (6.15) follows from equation
(6.14) or equivalently equation (6.20) and so the superconformal
transformations are encoded in $F_{B\dot B}$ subject to equation
(6.20).
 We shall refer to equation (6.20) as the superconformal
Killing equation, and the field
$F_{B\dot B}$ as the superconformal Killing vector, these being the
natural analogues of the conformal Killing
equation and the usual Killing vector in Minkowski space.

To find the consequences for the $x$-space component fields within
$F_{B\dot B}$ we  expand the superfield
$F_{B\dot B}$ as a Taylor series in $\theta$ and solve the
superconformal Killing  equation
order by order in $\theta$. Writing $F_{B\dot B}$  as
$$\eqalign{
F_{B\dot B}(x,\theta,\bar\theta)=&\zeta_{B\dot B}(x)+\theta^A
\chi_{AB\dot B}(x)+\bar\theta^{\dot A}\bar\chi_{\dot AB\dot
B}(x)+\theta^A\bar\theta^{\dot A}A_{AB\dot B\dot A}(x)\cr &+{1\over
2}\theta^2 f_{B\dot B}(x)+{1\over 2}\bar\theta^2g_{B\dot B}(x)\cr
&+{1\over 2}\theta^2\bar\theta^{\dot A}\lambda_{B\dot B\dot
A}(x)+{1\over 2}\bar\theta^2\theta^A\mu_{BA\dot B}(x)+{1\over
4}\theta^2\bar\theta^2\kappa_{B\dot B}(x)\cr }
\eqno(6.22)$$
and substituting this expression into the superconformal Killing
equation we find that  the resulting constraints are solved by the
solution
$$\eqalign{F_{B\dot B}&=\zeta_{B\dot B}
+\theta_B\bar\chi_{\dot B}-\bar\theta_{\dot
B}\chi_B+i\theta^A\bar\theta^{\dot A}\sigma^\mu_{B\dot
A}\partial_\mu\zeta_{A\dot B}\cr &+i\theta_B\bar\theta_{\dot
B}\alpha-{i\over 2}\theta^2\bar\theta^{\dot A}\sigma^\mu_{B\dot
B}\partial_\mu\bar\chi_{\dot B}\cr &+{i\over
2}\bar\theta^2\theta^A\sigma^\mu_{A\dot B}\partial_\mu\chi_B\cr}
\eqno(6.23)$$
In this equation  $\alpha$ is constant,  $\zeta_{B\dot B}$ is a
conformal Killing vector which satisfies
$$
\partial_{(A}^{\phantom{(A}(\dot A}\zeta_{B)}^{\phantom{B)}\dot
B)}=0
\eqno(6.24)
$$
and $\chi_B$ and
$\bar\chi_{\dot B}$ are conformal spinors which obey the relation
$$
\partial_{(A}^{\phantom{(A}\dot A}\chi_{B)} =0,\ \
\partial_{(\dot A}^{\phantom{(A}A}\bar\chi_{\dot
B)}=0\eqno(6.25)$$

The solutions to equations (6.24) and  (6.25) are given by
$$\eqalign{
\zeta_\mu&=a_\mu+\lambda x_\mu+\Lambda_\mu^{\phantom{\mu}\nu}x_\nu
+k_\mu x^2-2(k\cdot x)x_\mu\cr
\chi_B&=4(i\varepsilon_B+\bar\eta^{\dot B}\sigma^\mu_{B\dot B}
x_\mu)\cr
 \bar\chi_{\dot B}&=-4(i\bar\varepsilon_{\dot
B}+\eta^B\sigma^\mu_{B\dot B}x_\mu)\cr}
\eqno(6.26)$$
where $a_\mu$, $\lambda$,
$\Lambda_{\mu\nu}$, $\varepsilon_B$  and $\eta_B$ are constant
parameters.

Combining equations (6.26) and (6.23)  it is clear that the
parameters $a_\mu$, $\lambda$, $\Lambda_{\mu\nu}$, $k_\mu$, $\alpha$,
$\varepsilon_B$ and $\eta_B$ are translations, dilations, Lorentz
rotations, special conformal transformations, chiral transformations,
chiral rotations, supersymmetry transformations and special
supersymmetry transformations respectively.
\par
Having found the superconformal transformations on superspace  we
now turn our attention to the transformations  of superfields under a
superconformal transformation. If $\varphi$ is a general superfield,
which may carry Lorentz indices, then, its transformation is
of the form
$$\delta\varphi(z)=\delta
z^\pi\partial_\pi\varphi(z)+J\varphi(z)
\eqno(6.27)$$
where $ J$ is
a  superfield which arises from  the non-trivial action of generators
from  the isotropy group acting on $\varphi$ at the origin of the
superspace. This factor is most  pedagogically worked out by
considering the superfields as induced representations. However, here
we content ourselves with the final result which for a general
superfield
$\varphi$ is given by
$$\eqalign{
\delta\varphi =&(F\cdot\partial)\varphi-{1\over 8i}\sigma_\mu^{A\dot
A} (\bar D_{\dot A}F^\mu D_A\varphi+D_AF^\mu\bar D_{\dot A}\varphi)\cr
&+\left\{{1\over 4}(\partial\cdot F)\Delta+{1\over 96i}\sigma^{\mu
A\dot A} ([D_A,\bar D_{\dot A}]F_\mu){\cal
A}-2(\partial^{[\mu}F^{\nu]})\sigma_{\mu\nu}\right.\cr &\left.+{1\over
8}\partial^\mu(D^A F_\mu){\cal S}_A+{1\over 8}\partial^\mu(\bar D^{\dot
A}F_\mu)\bar{\cal S}_{\dot A}-{1\over 8}\partial^\mu(\partial\cdot
F)\kappa_\mu\right\}\varphi\cr }
\eqno(6.28)$$
In this equation the symbols $\{\Delta,
\Sigma_{\mu\nu}, {\cal A}, {\cal S}_A, \bar{\cal S}_{\dot A},
\kappa_\mu\}$ are constants that are the values of the
corresponding generators of the  isotropy group acting on  the
 superfield when it is taken to be at the origin of superspace.  For
almost all known situations,  only the
 parameters  $\Delta$, $\Sigma_{\mu \nu}$
and $\cal A$,  which correspond to the dilation,
Lorentz and
$U(1)$ transformations respectively, are non-zero.
The first part of
the result is just the shift in the coordinates  which is given by
$$\delta z^\pi\partial_\pi\varphi=(F\cdot\partial)\varphi-{1\over 8i}
\sigma_\mu^{A\dot A}(\bar D_{\dot A}F^\mu D_A\varphi+D_A
F^\mu\bar D_{\dot
A}\varphi)
\eqno(6.29)$$
while  $J$ is given by
$$\eqalign{
 J=&{1\over 4}(\partial\cdot F)\Delta+{1\over 96i}
\sigma^{\mu A\dot A}[D_A,\bar D_{\dot A}]F_\mu{\cal
A}-2\partial^{[\mu}F^{\nu]}
\sigma_{\mu\nu}\cr
&+{1\over 8}\partial^\mu(D^A F_\mu){\cal S}_A+{1\over
8}\partial^\mu(\bar D^{\dot A}F_\mu)\bar{\cal S}_{\dot A}-{1\over
8}\partial^\mu(\partial\cdot F)\kappa_\mu\cr }
\eqno(6.30)$$
\par
We can verify that equation (6.28) reproduces some of the known
results. Let us consider dilations which
are generated by taking $F_\mu=\lambda x_\mu$. For this case,
equation (6.28) becomes
 $$\delta\varphi=\lambda(x^\mu\partial_\mu+{1\over
2}\theta^A\partial_A+{1\over 2}
\bar\theta^{\dot A}\bar\partial_{\dot
A})\varphi+(\lambda)\Delta\varphi
\eqno(6.31)
$$
which  we recognise as the well known result.
In fact, by writing  $J$ as the most general form possible  which is
linear  in
$F_{B\dot B}$, contains covariant derivatives, is consistent
with dimensional analysis  and then evaluating the result for particular
transformations we can also arrive a the correct $J$.
\par
 We can apply equation (6.28) to the case of a chiral and
anti-chiral superfield. For simplicity, let us consider a lorentz
scalar chiral superfield whose ${\cal S}_A, \bar{\cal S}_{\dot A},
\kappa_\mu$ values also vanish. The result is
$$\delta\varphi=(F\cdot\partial)\varphi-{1\over 8i}\sigma_\mu^{A\dot A}
\bar D_{\dot A}F^\mu D_A\varphi
+\Delta ({1\over 4}(\partial\cdot F)\varphi+{1\over 48i}\sigma^{\mu
A\dot A}([D_A,\bar D_{\dot A}]F_\mu)\varphi)
\eqno(6.32)$$
and
$$\delta\bar\varphi=(F\cdot\partial)\bar\varphi-{1\over
8i}\sigma_\mu^{A\dot A}D_AF^\mu\bar D_{\dot A}\bar\varphi+ \Delta
({1\over 4}(\partial\cdot F)\bar\varphi-{1\over 48i}\sigma^{\mu A\dot
A}([D_A,\bar D_{\dot A}]F_\mu)\bar\varphi)
\eqno(6.33)
$$
The reader will observe that the dilation and $A$ weights of the
chiral superfield are tied together, a fact that can be established by
taking
the straightforward reduction of equation (6.28) and making sure the
transformed superfield is still   chiral or anti-chiral as
appropriate. We will discuss this result from a more general
perspective in the next section.
\bigskip
{\bf 6.2 Anomalous Dimensions of Chiral Operators at a Fixed Point}
\medskip
Let us consider a supersymmetric theory at a fixed point of the
renormalisation group, i.e. $\beta=0$. Such a theory should be
invariant under superconformal transformations. As in all
supersymmetric theories some of the observables are given by chiral
operators which by definition obey the equation
$$D_{\dot A} \varphi=0
\eqno(6.34)$$
where $\varphi$ denotes the chiral operator involved.
It follows that this equation must itself be invariant under any
superconformal transformation i.e. $D_{\dot A} \delta \varphi=0$.
Choosing a special supersymmetry transformation we conclude that
$$ \{D_{\dot A},\ S_{\dot B}\} \varphi= 0
\eqno(6.35)
$$
In this equation we can swop the covariant derivative for the
generator of supersymmetry transformations using the equation
$$ D_{\dot A}={\partial \over \partial \theta ^{\dot A}}-i(\sigma
^m)_{A{\dot  A}} \theta^{A}\partial_m = Q_{\dot A}
-2i(\sigma
^m)_{A{\dot  A}}\theta ^A \partial _m
\eqno(6.36)$$
We then conclude that
$$ \{Q_{\dot A},\ S_{\dot B}\}\varphi = 0
\eqno(6.37)
$$
plus terms that contain   space-time derivatives.
However, in this equation the condition must hold separately on the
parts of the equation containing space-time derivatives and those
that do not.
The advantage of writing the equation in this form is that
 the anti-commutator is one of the defining relations  of  the
superconformal algebra, namely
$$\{ Q_{\dot A},\ S_{\dot B}\} = \epsilon _{\dot A\dot B}(2D-4iA) +
 {(\bar {\sigma}^{\mu\nu})} _{\dot A\dot B}J_{\mu\nu}
\eqno(6.38)
$$
where $D$ and $A$ are the generators of the dilations and
U(1) transformations in the superconformal algebra which we gave this
algebra in lecture one.  If we restrict the superconformal algebra
to just its
super-Poincare subgroup then the $A$ generator is identified with the
generator of
$R$ transformations. The latter satisfies the relation
 $[Q_A,\ R]= iQ_A$ comparing this with the equivalent  commutator in
the  superconformal group (i.e. $[Q_A,\ A]= -i{3\over 4}Q_A$) we thus
find that the generators are related by $A= -{3\over 4}R$.

Consequently for a
Lorentz invariant chiral operator we conclude that
$D=-i{3\over 2}R$. One can also find this result by
substituting the explicit
expressions for $D_{\dot A}$ and $S_{\dot A}$ in equation (6.35)
and setting $\theta =0$.
We summarise the result in the theorem
\par
{\bf Theorem }[309]
\par
 Any Lorentz invariant operator in a four dimensional supersymmetric
theory at a fixed point has its anomalous dimension $\Delta$ and
 chiral $R$ weight,   related by the equation
 $$\Delta =-i{3\over 2}R. \eqno(6.39)
$$
\par
In any conformal theory we can determine the two and three
point Green's functions using conformal invariance alone.
However, one can not normally use this symmetry alone to fix the
anomalous weights of any operators. Since non-trivial fixed points
are outside the range of usual perturbation theory,  these must be
calculated using techniques such as the $\epsilon$-expansion. The
result so obtained are approximations and in some case one can not
reliably calculate the anomalous dimensions at all. However,
in supersymmetric theories at a fixed point one can determine
the anomalous dimensions of chiral operators in superconformal theories
exactly in terms of their $R$ weight. However, in many situations one
does know the
$R$ weight of the chiral operators of interest and we so can indeed
exploit the above theorem to find their anomalous dimensions
exactly [309]. We shall shortly demonstrate this procedure
with some examples.
\par
We must first fix the normalisation of the dilation and $R$ weights
that is implied by the superconformal algebra. The relation
$[P_\mu,\ D]=P_\mu$ implies that $P_\mu$ has dilation weight one.
On the other hand, the relationship
$[Q_A,\ R]= iQ_A$ implies that $Q_A$
has $R$ weight 1. Consequently, $\theta ^A$ has $R$ weight $-1$
meaning that it transforms as $\theta ^A \to e^{-i\alpha } \theta ^A$
where $\alpha$ is the parameter of $R$ transformations.
\par
As our first example, let us consider the Wess-Zumino model in four
dimensions and suppose that it had a non-trivial fixed point at which
the interaction was of the usual form;
$$\int d^4x d^2\theta \varphi^3
\eqno(6.40)$$
Using the above scaling of $\theta $ we find that $\varphi$ transforms
as $\varphi \to e^{i{2\over 3}\alpha } \varphi$ and as a
result $\varphi $ has $R$ weight ${2\over 3}$. Using our theorem we
find that
$\varphi$ had dilation weight one. This is the canonical dilation
weight of $\varphi$, that is, the weight it would have in the free
theory. It can be argued that if $\varphi$ has its canonical weight
then the theory must be free and so such a non-trivial fixed point
can not exist [310].  It can also be argued that this result implies
the the Wess-Zumino model is a trivial field theory meaning that the
only consistent value of the coupling constant as we remove the
cutoff is zero [311].
\par
 Now let us consider the Wess-Zumino model in three dimensions and
suppose it has a non-trivial fixed point at which the interaction
is given by
$$\int d^3x d^2\theta \varphi^3.
\eqno(6.41)$$
This is the supersymmetric generalisation of the Ising model.
Running through the same argument as above, but taking into account
the modified form of the three dimensional superconformal algebra, we
find that
$\varphi$ has anomalous  dimension ${1\over 6}$. The scaling
weight of a quantum operator is made up of the sum of its
canonical weight and its anomalous weight. The operator $\varphi$ has
canonical weight
$1/2$ and hence its scaling weight is ${2\over 3}$. Such a
non-trivial fixed point is known to exist by using the  epsilon
expansion which also gives an anomalous dimension in agreement with
this result [312]. Since the anomalous dimension is non-zero the
theory can not be a trivial theory, as is the case for
the Wess-Zumino model in four dimensions.
\par
We now briefly summarise the alternative argument given in reference
[312]  for calculating the anomalous dimensions of chiral operators
using the epsilon expansion. Let us consider a $d$ dimensional
theory whose action is given by
$$\int d^d x d^4 \theta \varphi_0\bar \varphi_0 +
(\int d^d x d^2 \theta g_0\varphi_0^n +c.c)
\eqno(6.42)
$$
where the action is given in terms of  bare quantities which  are
 denoted with a subscript $0$.
The dimension of $\varphi_0$ and $g_0$ are readily found to
be
${(d-2)\over 2}$ and
$d-1-{n\over 2}(d-2)$ respectively. The critical dimension, $d_c$ of
the theory is the  one where the coupling $g_0$ is dimensionless
and so is given by
$d_c=2{(n-1)\over (n-2)}$.
To renormalize the theory we introduce the wavefunction and coupling
renoramalization constants $Z$ and $Z_g$ respectively which relate
the  bare quantities  to the renormalized
quantities. The latter  are denoted by the same symbols, but
without a subscript $0$. The relationships between the bare and
renormalized quantities are given by the equations
$$\varphi_0= Z^{1\over 2}\varphi,\ \  {\rm {and}} \ \
g_0=\mu^{\epsilon {(n-2)\over 2}}Z_g g
\eqno(6.43)$$
where the constant $\mu$ is the
renormalization scale. It is raised to the above power in order
that $g$ be dimensionless.
\par
In the epsilon expansion method one carries
out a double perturbation expansion in $\epsilon\equiv d_c-d$ and $g$.
The beta-function, $\beta$ and anomalous
dimension  of $\varphi$, $\gamma$ are defined, as usual, by
$$\beta =\mu{\partial \over \partial \mu}g,\ \
\gamma=\mu{\partial \over \partial \mu}lnZ
\eqno(6.44)$$
where the differentiation is carried out for $g_0$ and the regulator,
$\Lambda$ held fixed. The non-renormalisation theorem implies that
$Z_gZ^{n\over 2}=1$ and as a result we find that
$$\beta = -{1\over 2}(n-2)\epsilon g+ng\gamma
\eqno(6.45)$$
Since we are interested in the anomalous dimension at a fixed point
(i.e. $\beta=0$) we must conclude that
$$\gamma= \epsilon{(n-2)\over 2n}={(n-1)\over n}-d{(n-2)\over 2n}
\eqno(6.46)$$
Using the renormalization group to calculate the
two point function of $\varphi$ we find that the actual scaling
dimension of $\varphi$ is the sum of its anomalous dimension and
its canonical dimension and so its scaling dimension  is the given by
$${(n-1)\over n}+{d\over n}-1
\eqno(6.47)$$
Substituting $d=n=3$ we recover the above result for the three
dimensional Ising model, while for the four dimensional Ising model
$\varphi$ has zero anomalous dimension. The calculation of the
anomalous dimensions of chiral operators in supersymmetric theories
is the only known case for which the  epsilon expansion gives
exact results. However, even in this case one must go further and
establish the existence of the non-trivial fixed point [312].
\par
The theorem in this section can also be used to fix the anomalous
dimensions for the chiral operators in the two dimensional $N=2$
supersymmetric Landau-Ginsburg models whose superpotential at
the fixed point take  the form
$$\int d^2 x d^2 \theta \varphi^n +c.c
\eqno(6.48)
$$
The anomalous dimensions  agree with the correspondence between
these models at their fixed points and the $N=2$ minimal series of
superconformal models. This result was first
conjectured in [317] and shown by using the epsilon expansion in
reference [312]. Using equation (6.45) we find that the anomalous
dimension of $\varphi$ is ${1\over n}$.
\par
The theorem can also be applied to four dimensional gauge invariant
operators  composed form  the $N=1$ Yang-Mills field strength $W_A$.
Such a connection was used to argue  that  $N=1$ super QED is
trivial [309] and has been used extensively by Seiberg in recent work
on  dualities between certain $N=1$ supersymmetric theories.
\bigskip
\centerline{\bf Acknowledgement}
I wish to thank World Scientific Publishing for their Kind
premission to reproduce some of the material from refernce [0] and
Neil Lambert for suggesting many  improvements.
\medskip
\bigskip
\centerline{\bf References}
\medskip
\item {[0]} P. West, "Introduction to Supersymmetry and Supergravity",
(1990),
 Extended and Revised Second Edition,  World Scientific Publishing,
Singapore.
\item {[1]} Y.A. Golfand and E.S. Likhtman, {\it JETP Lett.}
{\bf13}, 323 (1971).
\item {[2]} D.V. Volkov and V.P. Akulov, {\it Pis'ma Zh. Eksp.
Teor. Fiz.} {\bf16}, 621 (1972);
{\it Phys. Lett.} {\bf46B}, 109 (1973).
\item {[3]} J. Wess and B. Zumino, {\it Nucl. Phys.} {\bf B70},
139 (1974).
\item {[4]} S. Coleman and J. Mandula, {\it Phys. Rev.} {\bf159},
1251 (1967).
\item {[5]} R. Hagg, J. Lopuszanski and M. Sohnius, {\it Nucl.
Phys.} {\bf B88}, 61 (1975).
\item {[6]} P. van Nieuwenhuizen and P. West, {\it Principles of
Supersymmetry and Supergravity},
forthcoming book to be published by Cambridge University Press.
\item {[7]} P. Ramond, {\it Phys. Rev.} {\bf D3}, 2415 (1971); A.
Neveu and J.H. Schwarz, {\it Nucl. Phys.} {\bf B31}, 86 (1971); {\it
Phys. Rev.} {\bf D4}, 1109 (1971); J.-L. Gervais and B. Sakita, {\it
Nucl. Phys.} {\bf B34}, 477, 632 (1971); F. Gliozzi, J. Scherk and D.I.
Olive, {\it Nucl. Phys.} {\bf B122}, 253 (1977).
\item {[8]} J. Wess and B. Zumino, {\it Nucl. Phys.} {\bf B78}, 1
(1974).
\item {[9]} For a discussion of the Noether procedure in the
context of supergravity, see:
S. Ferrara, D.Z. Freedman and P. van Nieuwenhuizen, {\it Phys. Rev.}
{\bf D13}, 3214 (1976).
\item {[10]} S. Ferrara and B. Zumino, {\it Nucl. Phys.} {\bf
B79}, 413 (1974); A. Salam  and J. Strathdee, {\it Phys. Rev.} {\bf
Dll}, 1521 (1975).
\item {[11]} A. Salam and J. Strathdee,
{\it Nucl. Phys.} {\bf B80}, 499 (1974); M. Gell-Mann and Y. Neeman,
(1974) unpublished; W. Nahm, {\it Nucl. Phys.} {\bf B135}, 149
(1978). For a review, see: D.Z. Freedman in {\it Recent Developments
in Gravitation}, Carg\`ese (1978), eds. M. Levy and S. Deser (Gordon
and Breach, New York, 1979); S.  Ferrara and C. Savoy in {\it
Supergravity '81}, eds. S. Ferrara and J. Taylor (Cambridge University
Press, Cambridge, 1982).
\item {[12]} E.P. Wigner, {\it Ann.
of Math.} {\bf 40}, 149 (1939).
\item {[13]} P. van Nieuwenhuizen, {\it Phys. Rep.} {\bf 68}, 189
(198 1).
\item {[14]} D. Freedman, P. van Nieuwenhuizen and S. Ferrara,
{\it Phys. Rev.} {\bf D13}, 3214 (1976); {\it Phys. Rev.} {\bf D14},
912 (1976).
\item {[15]} S. Deser and B. Zumino, {\it Phys. Lett.} {\bf 62B},
335 (1976).
\item {[16]} K. Stelle and P. West, {\it Phys. Lett.} {\bf B74},
330 (1978).
\item {[17]} S. Ferrara and P. van Nieuwenhuizen, {\it Phys.
Lett.} {\bf B74}, 333 (1978).
\item {[18]} A. Chamseddine and P. West, {\it Nucl. Phys.} {\bf
B129}, 39 (1977).
\item {[19]} P. Townsend and P. van Nieuwenhuizen, {\it Phys.
Lett.} {\bf B67}, 439 (1977).
\item {[20]} J. Wess and B. Zumino, {\it Nucl. Phys.} {\bf B78}, 1
(1974).
\item {[21]} S. Ferrara in {\it Proceedings of the 9th
International Conference on General Relativity and Gravitation}
(1980), ed. Ernst Schmutzer.
\item {[22]} M. Sohnius, K. Stelle and P. West, in {\it Superspace
and Supergravity}, eds. S.W. Hawking and M. Rocek (Cambridge University
Press, Cambridge, 1981).
\item {[23]} A. Salam and J. Strathdee, {\it Phys. Lett.}
{\bf51B}, 353 (1974); P. Fayet, {\it Nucl. Phys.} {\bf B113}, 135
(1976).
\item {[24]} M. Sohnius, K. Stelle and P. West, {\it
Nucl. Phys.} {\bf B17}, 727 (1980); {\it Phys. Lett.} {\bf 92B}, 123
(1980).
\item {[25]} P. Fayet, {\it Nucl. Phys.} {\bf B113},
135 (1976).
\item {[26]} P. Breitenlohner and M. Sohnius, {\it Nucl. Phys.}
{\bf B178}, 151 (1981); M. Sohnius, K. Stelle and P. West, in {\it
Superspace and Supergravity}, eds. S.W. Hawking and M. Rocek
(Cambridge University Press, Cambridge, 1981). \item
{[27]} P. Howe and P. West, {\it "N=1, d=6 Harmonic Superspace"},
in preparation.
\item {[28]} G. Sierra and P.K. Townsend, {\it Nucl. Phys.} {\bf
B233}, 289 (1984); L. Mezincescu and Y.P. Yao, {\it Nucl. Phys.} {\bf
B241}, 605 (1984).
\item {[29]} F. Gliozzi, J. Scherk and D. Olive, {\it Nucl. Phys.}
{\bf B122}, 253 (1977); L. Brink, J. Schwarz and J. Scherk, {\it Nucl.
Phys.} {\bf B121}, 77 (1977).
\item {[30]} In this context, see: M. Rocek and W. Siegel, {\it
Phys. Lett.} {\bf 105B}, 275 (1981); V.0. Rivelles and J.G. Taylor,
{\it J. Phys. A. Math. Gen.} {\bf15}, 163 (1982).
\item {[31]} S. Ferrara, J. Scherk and P. van Nieuwenhuizen, {\it
Phys. Rev. Lett.} {\bf 37}, 1035 (1976); S. Ferrara, F. Gliozzi, J.
Scherk and P. van Nieuwenhuizen, {\it Nucl. Phys.} {\bf B117}, 333
(1976); P. Breitenlohner, S. Ferrara, D.Z. Freedman, F. Gliozzi, J.
Scherk and P. van Nieuwenhuizen, {\it Phys. Rev.} {\bf D15}, 1013
(1977); D.Z. Freedman, {\it Phys. Rev.} {\bf D15}, 1173 (1977).
\item {[32]} S. Ferrara and P. van Nieuwenhuizen, {\it Phys.
Lett.} {\bf 76B}, 404 (1978).
\item {[33]} K.S. Stelle and P. West, {\it Phys. Lett.} {\bf 77B},
376 (1978).
\item {[34]} S. Ferrara and P. van Nieuwenhuizen, {\it Phys.
Lett.} {\bf 78B}, 573 (1978).
\item {[35]} K.S. Stelle and P. West, {\it Nucl. Phys.} {\bf
B145}, 175 (1978).
\item {[36]} M. Sohnius and P. West, {\it Nucl. Phys.} {\bf B203},
179 (1982).
\item {[37]} R. Barbieri, S. Ferrara, D. Nanopoulos and K. Stelle,
{\it Phys. Lett.} {\bf 113B}, 219 (1982).
\item {[38]} E. Cremmer, S. Ferrara, B. Julia, J. Scherk and L.
Girardello, {\it Phys. Lett.} {\bf 76B}, 231 (1978).
\item {[39]} E. Cremmer, B. Julia, J. Scherk, S. Ferrara, L.
Girardello and P. van Nieuwenhuizen, {\it Nucl. Phys.} {\bf B147}, 105
(1979).
\item {[40]} E. Cremmer, S. Ferrara, L. Girardello and A. Van
Proeyen, {\it Nucl. Phys.} {\bf B212}, 413 (1983); {\it Phys. Lett.}
{\bf 116B}, 231 (1982).
\item {[41]} S. Deser, J. Kay and K. Stelle, Phys. Rev. Lett. 38,
527 (1977); S. Ferrara and B.  Zumino, {\it Nucl. Phys.} {\bf B134},
301 (1978).
\item {[42]} M. Sohnius and P. West, {\it Nucl. Phys.} {\bf B198},
493 (1982).
\item {[43]} S. Ferrara, L. Girardello, T. Kugo and A. Van
Proeyen, {\it Nucl. Phys.} {\bf B223}, 191 (1983).
\item {[44]} S. Ferrara, M. Grisaru and P. van Nieuwenhuizen, {\it
Nucl. Phys.} {\bf B138}, 430 (1978).
\item {[45]} B. de Wit, J.W. van Holten and A. Van Proeyen, {\it
Nucl. Phys.} {\bf B184}, 77 (1981); {\it Phys. Lett.} {\bf 95B}, 51
(1980); {\it Nucl. Phys.} {\bf B167}, 186 (1980).
\item {[46]} A. Salam and J. Strathdee, {\it Phys. Rev.} {\bf
Dll}, 1521 (1975); {\it Nucl. Phys.}{\bf B86}, 142 (1975).
\item {[47]} W. Siegel, {\it Phys. Lett.} {\bf 85B}, 333 (1979).
\item {[48]} R. Arnowitt and P. Nath, {\it Phys. Lett.} {\bf 56B},
117 (1975); L. Brink, M. Gell-Mann, P. Ramond and J. Schwarz, {\it
Phys. Lett.} {\bf 74B}, 336 (1978); {\bf 76B}, 417 (1978);
S. Ferrara and P. van Nieuwenhuizen, {\it Ann. Phys.} {\bf126}, 111
(1980); P. van Nieuwenhuizen and P. West, {\it Nucl. Phys.} {\bf
B169}, 501 (1980).
\item {[49]} M. Sohnius, {\it Nucl. Phys.} {\bf B165}, 483 (1980).
\item {[50]} P. Howe, K. Stelle and P. Townsend, {\it Nucl. Phys.}
{\bf B214}, 519 (1983).
\item {[51]} M. Grisaru, M. Rocek and W. Siegel, {\it Nucl. Phys.}
{\bf B159}, 429 (1979).
\item {[52]} E. Berezin, {\it The Method of Second Quantization}
(Academic Press, New York, 1960).
\item {[53]} A. Salam and J. Strathdee, {\it Nucl. Phys.} {\bf
B76}, 477 (1974); S. Ferrara, J. Wess and B. Zumino, {\it Phys. Lett.}
{\bf 51B}, 239 (1974).
\item {[54]} M. Grisaru, M. Rocek and W. Siegel, {\it Nucl. Phys.}
{\bf B159}, 429 (1979).
\item {[55]} J. Wess, Lecture Notes in Physics 77 (Springer,
Berlin, 1978).
\item {[56]} J. Gates and W. Siegel, {\it Nucl. Phys.} {\bf B147},
77 (1979).
\item {[57]} J. Gates, K. Stelle and P. West, {\it Nucl. Phys.}
{\bf B169}, 347 (1980).
\item {[58]} R. Grimm, M. Sohnius and J. Wess, {\it Nucl. Phys.}
{\bf B133}, 275 (1978).
\item {[59]} P. Breitenlohner and M. Sohnius, {\it Nucl. Phys.}
{\bf B178}, 151 (1981).
\item {[60]} P. Howe, K. Stelle and P. West, {\it Phys. Lett.}
{\bf 124B}, 55 (1983).
\item {[61]} P. Howe, K. Stelle and P. West, "$N=1, \ d=6$ Harmonic
Superspace", Kings College preprint.
\item {[62]} M. Sohnius, K. Stelle and P. West, in {\it Superspace
and Supergravity}, eds. S.W. Hawking and M. Rocek (Cambridge University
Press, Cambridge, 1981).
\item {[63]} A. Galperin, E. Ivanov, S. Kalitzin, V. Ogievetsky
and E. Sokatchev, Trieste preprint.
\item {[64]} L. Mezincescu, JINR report P2-12572 (1979).
\item {[65]} J. Koller, {\it Nucl. Phys.} {\bf B222}, 319
(1983); {\it Phys. Lett.} {\bf 124B}, 324 (1983).
\item {[66]} P. Howe, K. Stelle and P.K. Townsend, {\it Nucl.
Phys.} {\bf B236}, 125 (1984).
\item {[67]} A. Salam and J. Strathdee, {\it Nucl. Phys.} {\bf
B80}, 499 (1974).
\item {[68]} S. Ferrara, J. Wess and B. Zumino, {\it Phys. Lett.}
{\bf 51B}, 239 (1974).
\item {[69]} J. Wess and B. Zumino, {\it Phys. Lett.} {\bf 66B},
361 (1977); V.P. Akulov, D.V. Volkov and V.A. Soroka, {\it JETP Lett.}
{\bf 22}, 187 (1975).
\item {[70]} R. Arnowitt, P. Nath and B. Zumino, {\it Phys. Lett.}
{\bf 56}, 81 (1975); P. Nath and R. Arnowitt, {\it Phys. Lett.} {\bf
56B}, 177 (1975); {\bf 78B}, 581 (1978).
\item {[71]} N. Dragon, {\bf Z. Phys.} {\bf C2}, 62 (1979).
\item {[72]} E.A. Ivanov and A.S. Sorin, {\it J. Phys. A.
Math. Gen} {\bf 13}, 1159 (1980).
\item {[73]} That some representations do not generalize to
supergravity was noticed in: M. Fischler, {\it Phys. Rev.} {\bf D20},
1842 (1979).
\item {[74]} P. Howe and R. Tucker, {\it Phys. Lett.} {\bf 80B},
138 (1978).
\item {[75]} P. Breitenlohner, {\it Phys. Lett.} {\bf 76B}, 49
(1977); {\bf 80B}, 217 (1979).
\item {[76]} W. Siegel, {\it Phys. Lett.} {\bf 80B}, 224 (1979).
\item {[77]} W. Siegel, "Supergravity Superfields Without a
Supermetric", Harvard preprint HUTP-771 A068, {\it Nucl. Phys.} {\bf
B142}, 301 (1978); S.J. Gates Jr. and W. Siegel, {\it Nucl. Phys.}
{\bf B147}, 77 (1979).
\item {[78]} See also in this context: V. Ogievetsky and E.
Sokatchev, {\it Phys. Lett.} {\bf 79B}, 222 (1978).
\item {[79]} R. Grimm, J. Wess and B. Zumino, {\it Nucl. Phys.}
{\bf B152}, 1255 (1979).
\item {[80]} These constraints were first given by: J. Wess and B.
Zumino, {\it Phys. Lett.} {\bf 66B}, 361 (1977).
\item {[81]} J. Wess and B. Zumino, {\it Phys. Lett.} {\bf 79B},
394 (1978).
\item {[82]} P. Howe and P. West, {\it Nucl. Phys.} {\bf B238}, 81
(1983).
\item {[83]} P. Howe, {\it Nucl. Phys.} {\bf B199}, 309 (1982).
\item {[84]} A. Salam and J. Strathdee, {\it Phys. Rev.} {\bf
Dll}, 1521 (1975).
\item {[85]} S. Ferrara and 0. Piguet, {\it Nucl. Phys.} {\bf
B93}, 261 (1975).
\item {[86]} J. Wess and B. Zumino, {\it Phys. Lett.} {\bf 49B}, 52
(1974).
\item {[87]} J. lliopoulos and B. Zumino, {\it Nucl. Phys.} {\bf
B76}, 310 (1974).
\item {[88]} S. Ferrara, J. lliopoulos and B. Zumino, {\it Nucl.
Phys.} {\bf B77}, 41 (1974).
\item {[89]} D.M. Capper, {\it Nuovo Cim.} {\bf 25A}, 259 (1975);
R. Delbourgo, {\it Nuovo  Cim.} {\bf 25A}, 646 (1975).
\item {[90]} P. West, {\it Nucl. Phys.} {\bf B106}, 219 (1976); D.
Capper and M. Ramon Medrano, {\it J. Phys.} {\bf 62}, 269 (1976); S.
Weinberg, {\it Phys. Lett.} {\bf 62B}, 111 (1976).
\item {[91]} M. Grisaru, M. Rocek and W. Siegel, {\it Nucl. Phys.}
{\bf B159}, 429 (1979).
\item {[92]} B.W. Lee in {\it Methods in Field Theory}, Les
Houches 1975, eds. R. Balian and J. Zinn-Justin (North Holland,
Amsterdam and World Scientific, Singapore, 1981).
\item {[93]} W. Siegel, {\it Phys. Lett.} {\bf 84B}, 193 (1979);
{\it 94B}, 37 (1980).
\item {[94]} L.V. Avdeev, G.V. Ghochia and A.A. Vladiminov, {\it
Phys. Lett.} {\bf 105B}, 272 (1981); L.V. Avdeev and A.A. Vladiminov,
{\it Nucl. Phys.} {\bf B219}, 262 (1983).
\item {[95]} D.M. Capper, D.R.T. Jones and P. van Nieuwenhuizen,
{\it Nucl. Phys.} {\bf B167}, 479 (1980).
\item {[96]} G. 't Hooft and M. Veltman, {\it Nucl. Phys.} {\bf
B44}, 189 (1972); C. Bollini and J. Giambiagi, {\it Nuovo Cim.} {\bf
12B}, 20 (1972); J. Ashmore, {\it  Nuovo Cim. Lett.} {\bf 4}, 37
(1972).
\item {[97]} P. Howe, A. Parkes and P. West, {\it Phys. Lett.}
{\bf 147B}, 409 (1984); {\it Phys. Lett.} {\bf 150B}, 149 (1985).
 \item {[98]} J.W. Juer and D. Storey, {\it Nucl. Phys.} {\bf
B216}, 185 (1983); O. Piguet and K. Sibold, {\it Nucl. Phys.} {\bf
B248}, 301 (1984).
\item {[99]} E. Witten, {\it Nucl. Phys.} {\bf B188}, 52
(1981).
\item {[100]} E. Witten, in "Unity of the Fundamental
Interactions", Proceedings of the 19th Erice School 1981, ed. A
Zichichi (Plenum Press, 1983).
\item {[101]} A. Parkes and P. West, {\it Phys.
Lett.} {\bf 138B}, 99 (1984).
\item {[102]} L. Mezincescu and D.T.R. Jones, {\it Phys. Lett.}
{\bf 136B}, 242, 293 (1984).
\item {[103]} L. Mezincescu and D.T.R. Jones, {\it Phys. Lett.}
{\bf 138B}, 293 (1984).
\item {[104]} P. West, {\it Phys. Lett.} {\bf 136B}, 371 (1984).
\item {[105]} A. Parkes and P. West, "Three-Loop Results in
Two-Loop Finite Supersymmetric Gauge Theories", {\it Nucl. Phys.} {\bf
B256}, 340 (1985); M.T. Grisaru, B. Milewski and D. Zanon, {\it Phys.
Lett.} {\bf 155B}, 357 (1985).
\item {[106]} B.S. deWitt, {\it Dynamical Theory of Groups and
Fields} (Gordon and Breach, New York, 1978). R. Kallosh, {\it Nucl.
Phys.} {\bf B78}, 293 (1974); M. Grisaru, P. van Nieuwenhuizen and
C.C. Wu, {\it Phys. Rev.} {\bf D12}, 3202 (1975); J. Honerkamp, {\it
Nucl. Phys.} {\bf B36}, 130 (1971); {\bf B48}, 269 (1972); G. 't Hooft
in {\it Proceedings of XII Winter School of Theoretical Physics} in
Karpacz; B.S. deWitt in {\it Quantum Gravity}, Vol. 2, eds. C.J. Isham,
R. Penrose and D.W. Sciama (Oxford University Press; London, 1980).
L. Abbott, {\it Nucl. Phys.} {\bf B185}, 189 (1981); D. Boulware,
"Gauge Dependence of the  Effective Action", University of Washington
preprint RLO 1388-822 (1980).
\item {[107]} K. Schaeffer, L. Abbot and M. Grisaru, Brandeis
preprint.
\item {[108]} M. Grisaru and W. Siegel, {\it Nucl. Phys.} {\bf
B201}, 292 (1982); M. Rocek and W. Siegel, {\it Superspace}
(Benjamin/Cummings, Reading, Mass., 1983).
\item {[109]} S. Ferrara and B. Zumino, {\it Nucl. Phys.} {\bf
B79}, 413 (1974).
\item {[110]} D.R.T. Jones, {\it Phys. Lett.} {\bf 72B}, 199
(1977); E. Poggio and H. Pendleton, {\it Phys. Lett.} {\bf 72B}, 200
(1977).
\item {[111]} O. Tarasov, A. Vladimirov, A. Yu, {\it Phys. Lett.}
{\bf 93B}, 429 (1980); M.T. Grisaru, M. Rocek and W. Siegel, {\it
Phys. Rev. Lett.} {\bf 45}, 1063 (1980); W.E. Caswell and D. Zanon,
{\it Nucl. Phys.} {\bf B182}, 125 (1981).
\item {[112]} M. Sohnius and P. West, {\it Phys. Lett.} {\bf
100B}, 245 (1981).
\item {[113]} S. Ferrara and B. Zumino, unpublished.
\item {[114]} M. Grisaru and W. Siegel, {\it Nucl. Phys.} {\bf
B201}, 292 (1982).
\item {[115]} P. Howe, K. Stelle and P. Townsend, {\it Nucl.
Phys.} {\bf B236}, 125 (1984).
\item {[116]} S. Mandelstam, Proc. 21st Int. Conf. on High Energy
Physics, eds. P. Petiau and J. Pomeuf, {\it J. Phys.} {\bf 12}, 331
(1982).
\item {[117]} L. Brink, O. Lindgren and B. Nilsson, {\it Nucl.
Phys.} {\bf B212}, 401 (1983); {\it Phys. Lett.} {\bf 123B}, 328
(1983).
\item {[118]} D. Freedman, private communication.
\item {[119]} D.R.T. Jones, {\it Nucl. Phys.} {\bf B87}, 127
(1975).
\item {[120]} P. Howe, K. Stelle and P. West, {\it Phys. Lett.}
{\bf 124B}, 55 (1983).
\item {[121]} P. West in {\it Proceedings of the Shelter Island II
Conference on Quantum Field Theory and Fundamental Problems of
Physics}, eds. R. Jackiw, N. Khuri, S. Weinberg and E. Witten, M.I.T.
Press (June 1983). This work contains
a general review of the finiteness properties of supersymmetric
theories and in particular an account of the anomalies argument
applicable to $N=2$ theories.
This latter work was performed in collaboration with P. Howe.
\item {[122]} P. Howe and P. West, "The Two-Loop $\beta$-Function
in Models in Extended Rigid Supersymmetry", King's preprint, {\it Nucl.
Phys.} {\bf B242}, 364 (1984).
\item {[123]} A.A. Slavnov, {\it Teor. Mat. Fig.} {\bf 13}, 1064
(1972); B. W. Lee and J. Zinn-Justin, {\it Phys. Rev.} {\bf D5}, 3121
(1972).
\item {[124]} D.J. Gross and F. Wilczek, {\it Phys. Rev.} {\bf
D8}, 3633 (1973).
\item {[125]} T. Curtright, {\it Phys. Lett.} {\bf 102B}, 17
(1981); P. West, Higher Derivative Regulation of Supersymmetric
Theories, CALTEC preprint, CALT-68-1226.
\item {[126]} S.L. Adler and W.A. Bardeen, {\it Phys. Rev.} {\bf
182}, 1517 (1969).
\item {[127]} A. Zee, {\it Phys. Rev. Lett.} {\bf 29}, 1198
(1972); J.H. Lowenstein and B. Schroer, {\it Phys. Rev.} {\bf D6},
1553 (1972); {\bf D7}, 1929 (1973).
\item {[128]} M. Grisaru and P. West, "Supersymmetry and the
Adler-Bardeen Theorem", {\it Nucl. Phys.} {\bf B254}, 249 (1985).
\item {[129]} P. Howe and P. West, unpublished.
\item {[130]} S. Ferrara and B. Zumino, {\it Nucl. Phys.} {\bf
B79}, 413 (1974).
\item {[131]} R. Slansky, {\it Phys. Rep.} {\bf 79}, 1 (1981); D.
Olive, "Relation   Between Grand Unified and Monopole Theories", Erice
(1981); F. G\"ursey in {\it 1st Workshop on Grand Unification}, eds.
P.H. Frampton, S.H. Glashow and A. Yildiz, (Math.  Sci. Press,
Brooklyn, Massachusetts, 1980); p. 39; Mehmet Koca, {\it Phys. Rev.}
{\bf D42}, 2636, (1981); D. Olive and P. West, {\it Nucl. Phys.} {\bf
B217}, 248 (1983).
\item {[132]} A. Parkes and P. West, {\it Phys. Lett.} {\bf 127B},
353 (1983).
\item {[133]} R. Barbieri, S. Ferrara, L. Maiani, F. Palumbo and
A. Savoy, {\it Phys. Lett.} {\bf 115B}, 212 (1982).
\item {[134]} L. Susskind, private communication.
\item {[135]} S.J. Gates, M.T. Grisaru, M. Rocek and W. Siegel,
{\it Superspace} (Benjamin/Cummins, Reading, 1983).
\item {[136]} L. Alvarez-Gaum\'e and D.Z. Freedman, {\it Phys.
Lett.} {\bf 94B}, 171  (1980); {\it Comm. Math. Phys.} {\bf 80}, 443
(1981).
\item {[137]} R. Kallosh in {\it Supergravity '81}, eds. S.
Ferrara and J.G. Taylor (Cambridge University Press, Cambridge, 1982).
\item {[138]} N. Marcus, and A. Sagnotti, {\it Phys. Lett.} {\bf
135B}, 85 (1984).
\item {[139]} A. Parkes and P. West, {\it Phys. Lett.} {\bf 122B},
365 (1983).
\item {[140]} A. Parkes and P. West, {\it Nucl. Phys.} {\bf B222},
269 (1983).
\item {[141]} A. Namazie, A. Salam and J. Strathdee, {\it Phys.
Rev.} {\bf D28}, 1481 (1983).
\item {[142]} J.J. Van der Bij and Y.-P. Yao, {\it Phys. Lett.}
{\bf 125B}, 171 (1983).
\item {[143]} S. Rajpoot, J.G. Taylor and M. Zaimi, {\it Phys.
Lett.} {\bf 127B}, 347 (1983).
\item {[144]} A. Parkes and P. West, {\it Phys. Lett.} {\bf 127B},
353 (1983).
\item {[145]} J.-M. Fr\`ere, L. Mezincescu and Y.-P. Yao, {\it
Phys. Rev.} {\bf D29}, 1196 (1984).
\item {[146]} L. Girardello and M. Grisaru, {\it Nucl. Phys.} {\bf
B194}, 55 (1982); O. Piquet, K. Sibold and M. Schweda, {\it Nucl.
Phys.} {\bf B174}, 183 (1980).
\item {[147]} L. O'Raifeartaigh, {\it Nucl. Phys.} {\bf B96}, 331
(1975); P. Fayet, {\it Phys. Lett.} {\bf 58B}, 67 (1975).
\item {[148]} S. Dimopoulos and S. Raby, {\it Nucl. Phys.} {\bf
B192}, 353 (1981).
\item {[149]} I. Iba\~nez and G.G. Ross, {\it Phys. Lett.} {\bf
105B}, 439 (1981); {\it Phys. Lett.} {\bf 110B}, 215 (1982); J. Ellis,
I. Iba\~nez and G.G. Ross, {\it Phys. Lett.} {\bf 113B}, 283 (1982).
\item {[150]} P. Fayet and J. Iliopoulos, {\it Phys. Lett.}
{\bf 51B}, 461 (1974).
\item {[301]} P. West, "A Comment on the Non-Renomalization Theorem in
Supersymmetric  Theories", Phys Lett B258, (1991) 369.
\item {[302]}  I. Jack and T. Jones and  P. West, "Not the
no-renomalization Theorem" , Phys Lett B258, (1991) 375.
\item {[303]} P. West, "Quantum Corrections to the Supersymmetric
Effective  Superpotential and Resulting Modifications of Patterns of
Symmetry Breaking"  Phys. Lett. B261 (1991) 396.
\item {[304]} M. Shifman and A. Vainshtein, Nucl. Phys. B359 (1991)
571.
\item {[305]} L. Dixon,  V. Kaplunovsky and J. Louis, Nucl. Phys.
{\bf B355} (1991) 649.
\item {[306]} M. Sohnius, PhD thesis, University of Karslsruhe,
(19976); {\sl Phys. Rep.} {\bf 128} (1985), 39.
\item {[307]} B. P. Conlong and P. West, in
B. P. Conlong, Ph. D. Thesis, University of London (1993).
\item {[308]} I. Buchbinder and S. Kuenko, " Ideas and Methods of
Supersymmetry and Supergravity",  (1995),Institute of Physics;
John Park, "$N=1$ Superconformal Symmetry in 4 Dimensions",
hep-th/9703191.
\item {[309]} B. P. Conlong and P. C. West, {\sl J. Phys.} {\bf A26},
 (1993), 3325. (The abstract of this paper was changed at the request
of  the  referee  and it is best ignored)
\item {[310]} S. Ferrara,  J. Iliopoulos and B. Zumino, Nucl. Phys.
B77, (1974), 413.
\item {[311]} J. Verbaarschot and P. West, "Renormalons in
Supersymmetric Theories"  International Journal of Mod. Phys.
A Vol. 6 No. 13 (1991) 2361.
\item {[312]} P. Howe and P. West, "N=2 Superconformal Models,
Landau-Ginsburg Hamiltonians and the  epsilon-expansion", Phys. Lett.
B223 377 (1989).
\item {[313]} P. Howe and P. West, "Chiral Correlators in
Landau-Ginsburg Theories and N=2  Super-conformal models",  Phys.
Lett. B227 397 (1989).
\item {[314]} P. Howe and P. West, "Fixed Points in Multi-field
Landau-Ginsburg Models"  Phys. Lett.  B244 270 (1990).
\item {[315]} P. West, {\it Nucl. Phys.} {\bf B106}, 219
(1976).
\item {[316]} M. Grisaru, M. Rocek and W. Siegel,
{\it Nucl. Phys.} {\bf B159}, 429 (1979).
\item {[317]} D. Kastor, E. Martinec and S. Shenker, "RG flow in $N=1$
Discrete Series", (1988), EFI preprint 88-31.
\item {[318]} G. Sierre and P.K. Townsend, in Proceedings of
"Supersymmetry and Supergravity 1983", ed B. Milewski, World
Scientific, 1983;
S. J. Gates Jr., Nucl. Phys. B238 (1984) 349.
\item {[319]} N. Seiberg, Phys. Lett. 206B (1988) 75.
\item {[320]} S. Adler, J. Collins and A. Duncan, Phys. Rev. D15,
(1977) 1712;
 J. Collins and A. Duncan and S. Joglekar, Phys. Rev. D16,
(1977) 438;
N.K.Nielsen, Nucl. Phys. B120 (1977) 212.
\item {[321]}A. Pickering and  P. West, Phys. Lett. 383B (1996) 54.
\item {[322]} B. de Wit, M.T. Grisaru and M. Rocek, hep-th/9601115
\item {[323]} P. Howe and P. West,  Nucl. Phys. B486 (1997) 425.
\item {[324]} B. de Wit and A. Van Proeyen, Nucl. Phys B245 (1985) 89.
\item {[325]} This follows a discussion between P. West and
A. Van Proyen.

\end